\documentclass[aps,twocolumn,floatfix,showpacs,amsmath,amsfonts]{revtex4-1}

\usepackage{graphicx}
\usepackage{bm}
\usepackage{units}
\usepackage{color}

\begin{document}

\title{Impact of the valence band structure of Cu$_{2}$O on excitonic spectra}

\author{Frank Schweiner}
\author{J\"org Main}
\author{Matthias Feldmaier}
\author{G\"unter Wunner}
\affiliation{Institut f\"ur Theoretische Physik 1, Universit\"at Stuttgart,
  70550 Stuttgart, Germany}
\author{Christoph Uihlein}
\affiliation{Experimentelle Physik 2, Technische Universit\"at Dortmund, 44221 Dortmund, Germany}
\date{\today}

\begin{abstract}
We present a method to calculate the excitonic spectra of all direct semiconductors 
with a complex valence band structure. 
The Schr\"odinger equation is solved using a complete basis set with Coulomb Sturmian functions.
This method also allows for the computation of oscillator strengths.
Here we apply this method
to investigate the impact of the valence band structure of cuprous oxide $\left(\mathrm{Cu_{2}O}\right)$
on the yellow exciton spectrum. 
Results differ from those of J.~Thewes~\emph{et al.}
[Phys. Rev. Lett. \textbf{115}, 027402 (2015)]; the differences are discussed and explained.
The difference between the second and third Luttinger parameter can be determined 
by comparisons with experiments, however, the evaluation of all three Luttinger parameters is not uniquely possible.
Our results are consistent with band structure calculations. 
Considering also a finite momentum $\hbar K$ of the center of mass, we show that the large
$K$-dependent line splitting observed for the $1S$ exciton state by G.~Dasbach~\emph{et al.} [Phys. Rev. Lett. \textbf{91}, 107401 (2003)]
is not related to an exchange interaction but rather to the complex valence band structure of $\mathrm{Cu_{2}O}$.
\end{abstract}


\maketitle

\section{Introduction}

Excitons are the quanta of the fundamental
optical excitations in both insulators and semiconductors in the visible
and ultraviolet spectrum of light. Being composed of 
an electron and a positivly charged hole, Wannier excitons can be treated 
within the so-called simple band model as an analog of 
the hydrogen atom~\cite{TOE_5,NM5_7,TOE,NM5_9}.

This simple band model assumes that both the valence band and the 
conduction band are parabolic, isotropic and nondegenerate. 
However, in all crystals with zinc-blende and diamond structure 
the valence band is degenerate at the center of 
the first Brillouin zone~\cite{17_17,17_17_11}.
Consequently, an interpretation of experimental spectra in terms 
of the hydrogen-like description of excitons is often not possible~\cite{35}.

This is also true for cuprous oxide $\left(\mathrm{Cu_{2}O}\right)$, which is
one of the most interesting semiconductors relating to excitons due to the large excitonic
binding energy of $R_{\mathrm{exc}}\approx86\,\mathrm{meV}$~\cite{80,74}.
Only after Altarelli, Baldereschi and Lipari had developed the theory of 
excitons in semiconductors with degenerate valence bands in the 1970s~\cite{17_17_18,17_17_26,7_11,17_17,17_15},
a controversy regarding the correct assignment of the exciton states for $\mathrm{Cu_{2}O}$
could be settled by Ch.\ Uihlein \emph{et al.} in 1981~\cite{7}, i.e., almost 30 years after
the experimental discovery of excitons in $\mathrm{Cu_{2}O}$
by Gross and Karryjew~\cite{GRE_4}.

Very recently, new attention has been drawn to the field of excitons by an experimental 
observation of the so-called yellow exciton series in $\mathrm{Cu_{2}O}$ 
up to a large principal quantum number of $n=25$~\cite{GRE}.
Besides a variety of new experimental and theoretical investigations 
on this topic~\cite{QC,74,75,50}, the complex valence band structure 
of $\mathrm{Cu_{2}O}$ has also moved into the focus once 
again~\cite{28,80}.

In this paper we present a method to solve the 
cubic Hamiltonian of excitons, which accounts 
for the complex valence band structure of most semiconductors.
We solve the corresponding Schr\"odinger equation 
in a complete basis including the Coulomb-Sturmian functions, which also
allows the direct calculation of oscillator strengths from the excitonic wave function 
and is not limited to certain quantum numbers as in previous works~\cite{28,17_17}.
Using this method we will reinvestigate the calculations of Ref.~\cite{28}
to discuss the values of the three Luttinger parameters of $\mathrm{Cu_{2}O}$.
Deviations from previous results are observed and discussed.
\textcolor{black}{However, our method is of general applicability for 
all direct semiconductors with a complex valence band structure, e.g.,
GaAs~\cite{17_15}, CuBr~\cite{7_13}, and other compounds~\cite{17_17_18}.
Only the values of the material parameters used have to be replaced.
The decisive advantage of our method is the fact that it can also be used 
for the theoretical investigation of exciton spectra in external magnetic and electric fields, where 
the effects of the complex valence band 
structure are much more evident~\cite{35} and where other methods with a 
restricted amount of quantum numbers~\cite{17_15,28} may be too imprecise or too complex due to 
the strong mixing of different exciton states.
An application will be presented in Ref.~\cite{125}.}

In this paper will also show that a finite momentum $\hbar K$ of the 
center of mass leads to terms in the Hamiltonian, 
which were initially assigned to the exchange interaction~\cite{8,9}.
These terms are of the correct order of magnitude to describe the $K$-dependent 
experimentally observed line splitting of the $1S$ exciton~\cite{8,9}.

The paper is organized as follows:
In Sec.~\ref{sec:Theory} we present the theory of 
excitons for the case of degenerate valence bands.
At first, we describe in Sec.~\ref{sub:VBS} the valence band structure 
and the yellow exciton series of $\mathrm{Cu_{2}O}$. In this section
we already discuss the impact of the band structure on the 
exciton series qualitatively.
Having discussed the Hamiltonian of the exciton in Sec.~\ref{sub:Hamiltonian},
we introduce our complete basis in Sec.~\ref{sub:Basis} and describe how to 
calculate oscillator strengths in Sec.~\ref{sub:Oscillator-strengths-1}.
We investigate the excitonic spectra in Sec.~\ref{sub:Determiantion-of}
and discuss the values of the three Luttinger parameters.
The treatment of the motion of the center of mass is presented 
in Sec.~\ref{sub:-dependent-line-splitting}.
Finally, we give a short summary and outlook in Sec.~\ref{sec:Summary-and-outlook}.

\section{Theory\label{sec:Theory}}

\subsection{Valence band structure and the yellow exciton series in Cu$_{2}$O \label{sub:VBS}}

Concerning its hydrogen-like spectrum up to a 
principal quantum number of $n=25$~\cite{GRE}, the yellow exciton 
in $\mathrm{Cu_{2}O}$ seems to be a 
perfect example of a Wannier exciton.
However, a more precise investigation of this 
spectrum shows clear deviations from the 
simple model with spherical effective masses~\cite{7,28}.
These deviations can be explained in terms of the complex valence band 
structure of $\mathrm{Cu_{2}O}$.

\begin{figure}
\begin{centering}
\includegraphics[width=0.8\columnwidth]{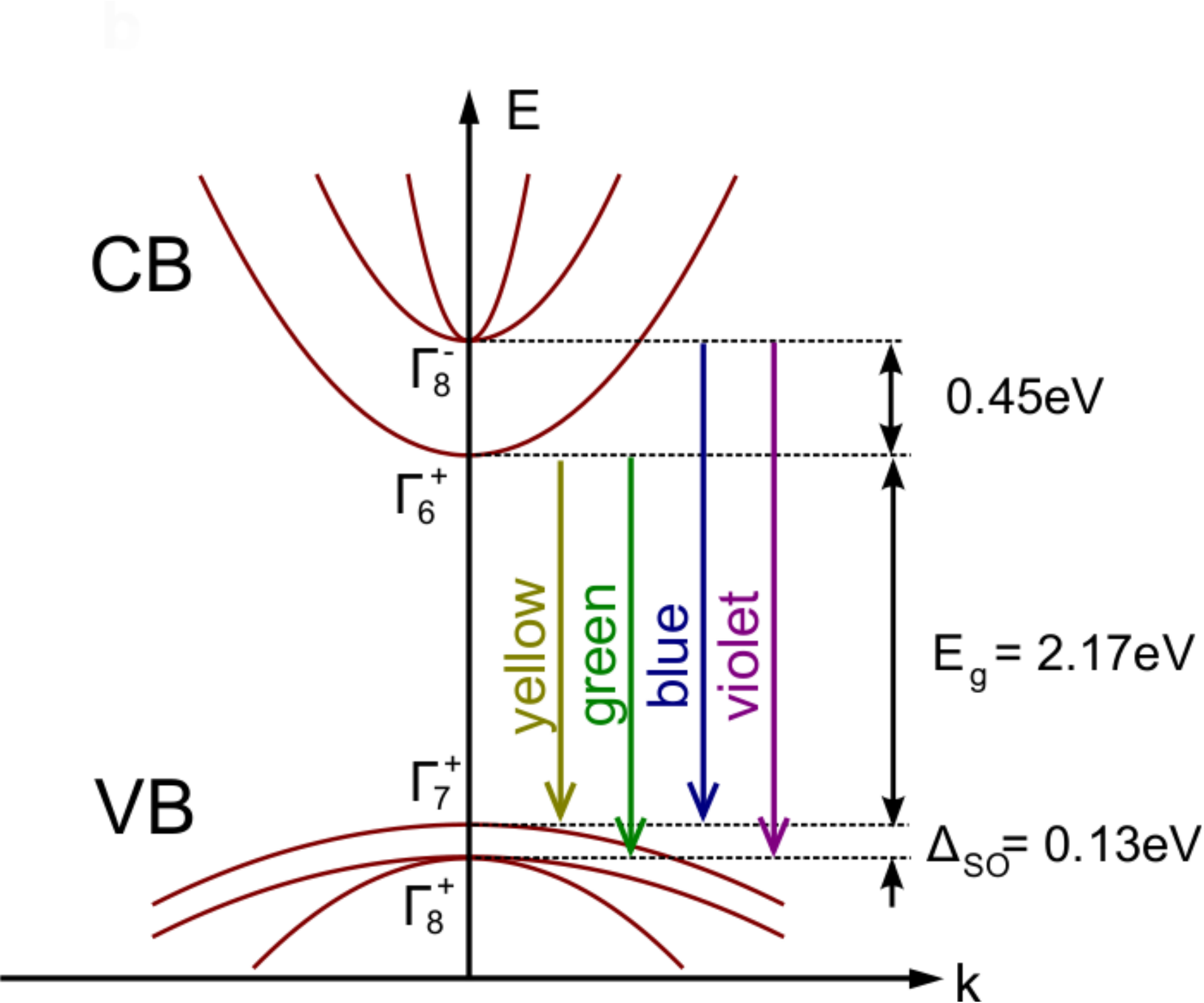}
\par\end{centering}

\protect\caption{(Color online) Band structure of $\mathrm{Cu_{2}O}$~\cite{GRE}. 
Due to the spin-orbit coupling
(\ref{eq:soc}) the valence band splits into a lower lying fourfold-degenerate
band $\left(\Gamma_{8}^{+}\right)$ of and a higher lying twofold-degenerate
band $\left(\Gamma_{7}^{+}\right)$. The lowest lying conduction band
has $\Gamma_{6}^{+}$ symmetry. Depending on the bands involved, one
distinguishes between the yellow, green, blue, and violet exciton series.
\label{fig:Band-structure-of}}

\end{figure}

Without spin-orbit coupling the valence
band in $\mathrm{Cu_{2}O}$ has the
symmetry $\Gamma_{5}^{+}$ and is threefold degenerate
at the $\Gamma$-point or the center
of the Brillouin zone. This degeneracy can be accounted for by a quasi-spin
$I=1$, \textcolor{black}{which is a convenient abstraction to denote the three 
orbital Bloch functions $xy$, $yz$, and $zx$, which transform according to $\Gamma_{5}^{+}$.}
Since $\mathrm{Cu_{2}O}$ has cubic symmetry, the symmetry of the
bands can be assigned by the irreducible representations $\Gamma_{i}^{\pm}$
of the cubic group $O_{\mathrm{h}}$, where the superscript $\pm$
denotes the parity. 
Considering the spin-orbit coupling between the quasi-spin $I$ and the spin 
$S_{\mathrm{h}}$ of a hole in the valence band,
this sixfold degenerate band (now including the hole spin)
splits into a lower lying fourfold-degenerate
band $\left(\Gamma_{8}^{+}\right)$ and a higher lying twofold-degenerate
band $\left(\Gamma_{7}^{+}\right)$ by an amount of $\Delta$, which
is the spin-orbit coupling constant (see Fig.~\ref{fig:Band-structure-of}).
The presence of the nonspherical symmetry of
the solid as well as interband interactions cause these bands
to be nonparabolic but deformed.

Neglecting these effects at first, one arrives at the
simple-band model and can distinguish between four exciton series 
depending on the valence band and the conduction band 
involved (see Fig.~\ref{fig:Band-structure-of}).
Within this model the wave function of an exciton
consists of the so-called envelope function, which describes 
the relative movement of the electron and the hole, and the Bloch functions of 
the bands involved~\cite{TOE}.

Due to the spins of electron and hole, e.g., the yellow 
exciton series is fourfold degenerate.
The presence of an exchange interaction between the spins of the electron and the hole
lifts this degeneracy and leads to ortho and para excitons~\cite{GRE,SO}.
While the threefold degenerate ortho excitons can be observed 
in absorption spectra, the nondegenerate para excitons
are spin-flip-forbidden~\cite{SO}.

Going now beyond the simple-band model, the anisotropic dispersion of 
the valence band has a significant influence 
on the excitons of the yellow series.
The anisotropic dispersion leads to a coupling between the relative 
motion of the electron and the hole and the orbital Bloch functions
$xy$, $yz$, and $zx$~\cite{G1} of the original
$\Gamma_{5}^{+}$-band. This will be described mathematically by the 
so-called $H_{\mathrm{d}}$-term in Sec.~\ref{sub:Basis}.

As can be seen from Fig.~\ref{fig:Band-structure-of}, the yellow exciton series is connected with the 
$\Gamma_{7}^{+}$-valence band. Due to symmetry considerations, 
the amplitudes, which describe the magnitude of the 
contribution of the orbital Bloch functions $xy$, $yz$ and $zx$ to this band, must 
have the same absolute value. Thus, the anisotropy of the 
$\Gamma_{5}^{+}$-Bloch functions is compensated in the $\Gamma_{7}^{+}$-band.
The same statement now also holds for all nondegenerate exciton states:
for reasons of symmetry, the wave function of para excitons contains the 
orbital Bloch functions with amplitudes having the same absolute value.
Thus, the dispersion of the para exciton can be described by an isotropic
exciton mass.

As regards the threefold degenerate ortho exciton, 
the situation is different: 
each of the three exciton states can have a 
larger contribution of one of the orbital Bloch functions, respectively, 
without a violation of symmetry.
The disparity in the orbital Bloch components of the ortho exciton 
is caused by an admixture of the $\Gamma_{8}^{+}$-valence band 
via the $H_{\mathrm{d}}$-term. This disparity has then an impact
on the relative motion of electron and hole.
As a consequence, the envelope function of the ortho exciton has no 
further spherical or cubic symmetry but $D_{\mathrm{4h}}$-symmetry. 
Since $D_{\mathrm{4h}}$ is a subgroup of $O_{\mathrm{h}}$, it contains all 
symmetry operations, which leave a given $\Gamma_{5}^{+}$-component 
of the ortho exciton invariant. For instance, in the case of the $xy$-component the symmetry 
axis of the according subgroup $D_{\mathrm{4h}}$ is the $z$-axis of the crystal.
So the reduction of the symmetry of the envelope function
reflects the anisotropic dispersion of the orbital Bloch functions.
Due to the $H_{\mathrm{d}}$-term the cubic symmetry 
$O_{\mathrm{h}}$ holds no further for 
the Bloch functions and the relative motion separately but only for the 
combined function.

The lower symmetry of the envelope function 
allows for a smaller mean distance between electron and hole 
in a specific direction,
which leads to a gain of energy due to the Coulomb interaction.
This effect may be compared to the Jahn-Teller effect, where a reduction of 
symmetry in connection with degeneracies leads to a 
gain of energy in the system. 

As can be seen, there is a close connection between degeneracy 
and symmetry reduction of the envelope function. This fact explains 
one of the most striking features of excitonic spectra in $\mathrm{Cu_{2}O}$: 
the visibility of $D$ and $F$ exciton states, i.e., 
exciton states with angular momentum $L=2$ and $L=3$~\cite{7,28}. The visibility arises due to 
the admixture of quadrupole or dipole-allowed $S$- and $P$-exciton states.

In the following sections~\ref{sub:Hamiltonian} and~\ref{sub:Basis} we will now 
introduce the problem of excitons in $\mathrm{Cu_{2}O}$
from a more mathematical point of view.

\subsection{Hamiltonian\label{sub:Hamiltonian}}

Via $\boldsymbol{k}\cdot\boldsymbol{p}$-perturbation theory and symmetry
considerations one can derive the Hamiltonian or the kinetic energy
of an electron within the valence band structure described in Sec.~\ref{sub:VBS}~\cite{SO,25,80}:
\begin{eqnarray}
H_{\mathrm{vb}}\left(\boldsymbol{k}\right) & = & -H_{\mathrm{so}}+\left(1/2m_0\right)\left\{\boldsymbol{k}^{2}\left[\hbar^{2}A_{1}+2B_{1}\left(\boldsymbol{I}\boldsymbol{S}_{\mathrm{h}}\right)\right]\right.\phantom{\frac{1}{1}}\nonumber \\
 &  & +A_{2}\left(k_{1}^{2}\left(\boldsymbol{I}_{1}^{2}-\boldsymbol{I}^{2}/3\right)+\mathrm{c.p.}\right)\phantom{\frac{1}{1}}\nonumber \\
 &  & +B_{2}\left(2 k_{1}^{2}\left(\boldsymbol{I}_{1}\boldsymbol{S}_{\mathrm{h}1}-\boldsymbol{I}\boldsymbol{S}_{\mathrm{h}}/3\right)+\mathrm{c.p.}\right)\phantom{\frac{1}{1}}\nonumber \\
 &  & +A_{3}\left(2\left\{ k_{1},\, k_{2}\right\}\left\{ \boldsymbol{I}_{1},\,\boldsymbol{I}_{2}\right\}+\mathrm{c.p.}\right)\phantom{\frac{1}{1}}\nonumber \\
 &  & +B_{3}\left(2\left\{ k_{1},\, k_{2}\right\}\left(\boldsymbol{I}_{1}\boldsymbol{S}_{\mathrm{h}2}+\boldsymbol{I}_{2}\boldsymbol{S}_{\mathrm{h}1}\right)+\mathrm{c.p.}\right)\left.\right\}\phantom{\frac{1}{1}}\label{eq:Hevb}
\end{eqnarray}
with $\left\{ a,b\right\} =\frac{1}{2}\left(ab+ba\right)$, the free
electron mass $m_{0}$, and c.p.\ denoting cyclic permutation. The spin-orbit
coupling reads~\cite{7,28}
\begin{equation}
H_{\mathrm{so}}=\frac{2}{3}\Delta\left(1+\frac{1}{\hbar^{2}}\boldsymbol{I}\boldsymbol{S}_{\mathrm{h}}\right).\label{eq:soc}
\end{equation}
Note that we use, in contrast to Ref.~\cite{80}, the
energy shift of $2\Delta/3$, by which the energy of the $\Gamma_{7}^{+}$-band
is set to zero at the $\Gamma$-point. 
Furthermore, we use the spin matrices of spin 1/2 for the hole spin
instead of the Pauli matrices.
The matrices of the quasi-spin $I=1$ are defined as in Ref.~\cite{25},
\begin{equation}
\boldsymbol{I}_{k}=\sum_{l,m}-i\hbar\varepsilon_{klm}\left(\hat{\boldsymbol{e}}_l\otimes\hat{\boldsymbol{e}}_m\right),
\label{eq:I}
\end{equation}
with the unit vectors $\hat{\boldsymbol{e}}_i$ and the Levi-Civita symbol $\varepsilon_{klm}$.

\begin{table}

\protect\caption{Material parameters used in the calculations. In Sec.~\ref{sub:Determiantion-of}
we use two different sets of the parameters $\Delta$, $\gamma_{1}'$, and $\mu'$.
For further information see text.\label{tab:1}}

\begin{centering}
\begin{tabular}{lll}
\hline 
\hline
Band gap energy & $E_{\mathrm{g}}=2.17028\,\mathrm{eV}$ & [16]\tabularnewline
Electron mass & $m_{\mathrm{e}}=0.99\, m_{0}$ & [34]\tabularnewline
Dielectric constant & $\varepsilon=7.5$ & [35]\tabularnewline
Exchange interaction & $J_0=12\,\mathrm{meV}$ & [8]\tabularnewline
Spin-orbit coupling & $\Delta=0.131\,\mathrm{eV}$ & [8]\tabularnewline
Valence band parameters & $\gamma_{1}'=2.77$ & [8]\tabularnewline
 & $\mu'=0.0586$ & [8]\tabularnewline
 & $\eta_1=-0.02$ & [8]\tabularnewline 
 & $\nu=2.167$ & [8]\tabularnewline
 & $\tau=1.5$ & [8]\tabularnewline
 & & \tabularnewline
Spin-orbit coupling & $\Delta=0.1338\,\mathrm{eV}$ & [14,20]\tabularnewline
Valence band parameters & $\gamma_{1}'=2.78$ & [14,20]\tabularnewline
 & $\mu'=0.47$ & [14,20]\tabularnewline
\hline 
\hline
\end{tabular}
\par\end{centering}

\end{table}

Very recently, the parameters $A_{i}$ and $B_{i}$ in 
Eq.~(\ref{eq:Hevb}) have been obtained~\cite{80}
by fitting the Hamiltonian to results of band structure
calculations~\cite{20}:
\begin{subequations}
\begin{eqnarray}
A_{1} & = & -1.76,\quad A_{2}=4.519,\quad A_{3}=-2.201,\\
B_{1} & = & 0.02,\quad B_{2}=-0.022,\quad B_{3}=-0.202.
\end{eqnarray}
\end{subequations}
In the case of an exciton one generally treats the missing electron
in the valence band as a hole, i.e., a quasi-particle with an energy
being opposite to the energy of the other electrons in the valence
band. Using the definition of the three Luttinger parameters,
\begin{subequations}
\begin{equation}
\gamma_{1}=-A_{1},\quad\gamma_{2}=\frac{1}{6}A_{2},\quad\gamma_{3}=\frac{1}{6}A_{3},
\end{equation}
and defining by analogy
\begin{equation}
\eta_{1}=-B_{1},\quad\eta_{2}=\frac{1}{6}B_{2},\quad\eta_{3}=\frac{1}{6}B_{3},
\end{equation}
\end{subequations}
the Hamiltonian of the hole reads~\cite{SO,SST,17_19,28} 
\begin{eqnarray}
H_{\mathrm{h}}\left(\boldsymbol{p}\right) & = & H_{\mathrm{so}}+\left(1/2\hbar^{2}m_{0}\right)\left\{ \hbar^{2}\left(\gamma_{1}+4\gamma_{2}\right)\boldsymbol{p}^{2}\right.\phantom{\frac{1}{1}}\nonumber \\
 &  & +2\left(\eta_{1}+2\eta_{2}\right)\boldsymbol{p}^{2}\left(\boldsymbol{I}\boldsymbol{S}_{\mathrm{h}}\right)\phantom{\frac{1}{1}}\nonumber \\
 &  & -6\gamma_{2}\left(p_{1}^{2}\boldsymbol{I}_{1}^{2}+\mathrm{c.p.}\right)-12\eta_{2}\left(p_{1}^{2}\boldsymbol{I}_{1}\boldsymbol{S}_{\mathrm{h}1}+\mathrm{c.p.}\right)\phantom{\frac{1}{1}}\nonumber \\
 &  & -12\gamma_{3}\left(\left\{ p_{1},p_{2}\right\} \left\{ \boldsymbol{I}_{1},\boldsymbol{I}_{2}\right\} +\mathrm{c.p.}\right)\phantom{\frac{1}{1}}\nonumber \\
 &  & \left.-12\eta_{3}\left(\left\{ p_{1},p_{2}\right\} \left(\boldsymbol{I}_{1}\boldsymbol{S}_{\mathrm{h}2}+\boldsymbol{I}_{2}\boldsymbol{S}_{\mathrm{h}1}\right)+\mathrm{c.p.}\right)\right\}. \phantom{\frac{1}{1}}\label{eq:Hh}
\end{eqnarray}
The parameters in Eq.~(\ref{eq:Hh}) describe the dispersion of the hole in the
vicinity of the $\Gamma$-point: $\gamma_{1}$ and $\eta_1$ determine
the average effective mass 
of the hole while the other parameters
describe the splitting of the bands in the vicinity
of the $\Gamma$-point and the so-called band warping or the nonspherical
symmetry of the bands~\cite{SO}. 

The Hamiltonian of the exciton is given by~\cite{17_17,7}
\begin{equation}
H=E_{\mathrm{g}}+H_{\mathrm{e}}\left(\boldsymbol{p}_{\mathrm{e}}\right)+H_{\mathrm{h}}\left(\boldsymbol{p}_{\mathrm{\mathrm{h}}}\right)+V+H_{\mathrm{exch}}+H_{\mathrm{C}}\label{eq:Hpeph}
\end{equation}
with the energy $E_{\mathrm{g}}$ of the band gap and the kinetic energy
of the electron, 
\begin{equation}
H_{\mathrm{e}}\left(\boldsymbol{p}_{\mathrm{e}}\right)=\frac{\boldsymbol{p}_{\mathrm{e}}^{2}}{2m_{\mathrm{e}}}.
\end{equation}
Here $m_{\mathrm{e}}$ denotes the effective mass of the electron. The
Coulomb interaction, which is screened by the dielectric constant
$\varepsilon$, reads
\begin{equation}
V\left(\boldsymbol{r}_{e}-\boldsymbol{r}_{h}\right)=-\frac{e^{2}}{4\pi\varepsilon_{0}\varepsilon}\frac{1}{\left|\boldsymbol{r}_{e}-\boldsymbol{r}_{h}\right|}.
\end{equation}
The last two terms of Eq.~(\ref{eq:Hpeph}) are given by~\cite{7,Dasbach_83,PRB23} 
\begin{equation}
H_{\mathrm{exch}}=J_{0}\left(\frac{1}{4}-\frac{1}{\hbar^{2}}\boldsymbol{S}_{\mathrm{e}}\boldsymbol{S}_{\mathrm{h}}\right)\delta\left(\boldsymbol{r}\right)
\end{equation}
and~\cite{1} 
\begin{equation}
H_{\mathrm{C}}=\frac{a^2}{24\hbar^2 }\left(\frac{c_{\mathrm{e}}}{m_{\mathrm{e}}}\boldsymbol{p}_{\mathrm{e}}^4
+\frac{c_{\mathrm{h}}}{m_{\mathrm{h}}}\boldsymbol{p}_{\mathrm{h}}^4\right)-d\frac{e^2 a^2}{\varepsilon_0 \varepsilon^2}\delta\left(\boldsymbol{r}_{e}-\boldsymbol{r}_{h}\right)\label{eq:CCC}
\end{equation}
and denote the exchange interaction as well as 
the central-cell corrections. 
The coefficients $c_{\mathrm{e}}=1.35$, $c_{\mathrm{h}}=1.35$, and $d=0.18$ were calculated in Ref.~\cite{1}
within the simple band model and with $m_{\mathrm{h}}=0.69\,m_0$ for the hole mass.
Here $a=4.26\times10^{-10}\,\mathrm{m}$ is the lattice constant.

In the case of the $1S$ exciton, i.e., the ground state of the 
exciton, the wave
function is highly localized in position space and
comprises only a few unit cells of the solid, for which reason terms
of the order four in the momentum have to be considered.
Besides the $\boldsymbol{p}^{4}$-terms in Eq.~(\ref{eq:CCC}), one could also imagine terms
of the form $p_1^4+p_2^4+p_3^4$, which have cubic symmetry.
The last term in Eq.~(\ref{eq:CCC}) appears due to 
corrections in the dielectric constant since
the continuum approach is not valid for the $1S$ exciton.

For subsequent calculations
it is appropriate to write the Hamiltonian~(\ref{eq:Hpeph}) in terms of irreducible
tensors~\cite{ED,7_11,44}, where we additionally set the position and the 
momentum of the center of mass to zero:
\begin{widetext}
\begin{eqnarray}
H & = & E_{\mathrm{g}}-\frac{e^{2}}{4\pi\varepsilon_{0}\varepsilon}\frac{1}{r}+\frac{2}{3}\Delta\left(1+\frac{1}{\hbar^{2}}I^{(1)}\cdot S_{\mathrm{h}}^{(1)}\right)+ H_{\mathrm{exch}}+H_{\mathrm{C}}\nonumber \\
\nonumber \\
 &  & +\frac{\gamma'_{1}}{2\hbar^{2}m_{0}}\left[\hbar^{2}p^{2}-\frac{\mu'}{3}P^{(2)}\cdot I^{(2)}+\frac{\delta'}{3}\left(\sum_{k=\pm4}\left[P^{(2)}\times I^{(2)}\right]_{k}^{(4)}+\frac{\sqrt{70}}{5}\left[P^{(2)}\times I^{(2)}\right]_{0}^{(4)}\right)\right]
\nonumber \\
 &  & +\frac{3\eta_{1}}{\hbar^{2}m_{0}}\left[\frac{1}{3}p^{2}\left(I^{(1)}\cdot S_{\mathrm{h}}^{(1)}\right)-\frac{\nu}{3}\, P^{(2)}\cdot D^{(2)}+\frac{\tau}{3}\,\left(\sum_{k=\pm4}\left[P^{(2)}\times D^{(2)}\right]_{k}^{(4)}+\frac{\sqrt{70}}{5}\left[P^{(2)}\times D^{(2)}\right]_{0}^{(4)}\right)\right]\label{eq:H0}
\end{eqnarray}
\end{widetext}
The first-order and second-order tensor operators correspond, as in
Ref.~\cite{44}, to the vector operators $\boldsymbol{r}$,~$\boldsymbol{I}$,~$\boldsymbol{S}_{\mathrm{e/h}}$,
and to the second-rank Cartesian operators 
\begin{subequations}
\begin{eqnarray}
I_{mn} & = & 3\left\{ I_{m},\, I_{n}\right\} -\delta_{mn}I^{2},\\
P_{mn} & = & 3\left\{ p_{m},\, p_{n}\right\} -\delta_{mn}p^{2},
\end{eqnarray}
\end{subequations}
respectively. We also use the abbreviation
\begin{equation}
D_{k}^{(2)}=\left[I^{(1)}\times S_{\mathrm{h}}^{(1)}\right]_{k}^{(2)}.
\end{equation}
The coefficients $\gamma'_{1}$, $\mu'$, and $\delta'$
are given by~\cite{7_11,7}
\begin{subequations}
\begin{equation}
\gamma'_{1}=\gamma_{1}+\frac{m_{0}}{m_{\mathrm{e}}},\quad\mu'=\frac{6\gamma_{3}+4\gamma_{2}}{5\gamma'_{1}},\quad\delta'=\frac{\gamma_{3}-\gamma_{2}}{\gamma'_{1}}
\end{equation}
and we define by analogy
\begin{equation}
\nu=\frac{6\eta_{3}+4\eta_{2}}{5\eta_{1}},\quad\tau=\frac{\eta_{3}-\eta_{2}}{\eta_{1}}.
\end{equation}
\end{subequations}

Since $\eta_i\ll\gamma_i$
holds in Eq.~(\ref{eq:Hh}), we neglect the corresponding terms of the 
Hamiltonian~(\ref{eq:H0}) in the following and use them only 
for the calculations at the end of Sec.~\ref{sub:Determiantion-of}.
The material parameters used in our calculations
are listed in Table~\ref{tab:1}. 

The parameters taken from Ref.~\cite{7}
have been obtained as fit parameters to excitonic spectra
using the spherical model, i.e., the model in which 
the $\delta'$\nobreakdash-dependent terms are neglected. 
Recent calculations on the band structure of 
$\mathrm{Cu_{2}O}$~\cite{20} yielded different values for 
the corresponding material parameters~\cite{80} 
showing that the spherical model by which $\mu'=0.47$
had been obtained may be inappropriate since $|\delta'|\gg|\mu'|$ holds.
These parameters are listed in Table~\ref{tab:1}, as well.

\subsection{Choice of the basis set\label{sub:Basis}}

To find an appropriate basis set to solve the Schr\"odinger equation, we
have to discuss the different terms of the Hamiltonian~(\ref{eq:H0})
as in Ref.~\cite{7}. 
The Hamiltonian 
\begin{equation}
H_{\mathrm{sb}}=E_{\mathrm{g}}-\frac{e^{2}}{4\pi\varepsilon_{0}\varepsilon}\frac{1}{r}+\frac{\gamma'_{1}}{2m_{0}}p^{2}+H_{\mathrm{C}}
\end{equation}
\\is the hydrogen-like Hamiltonian of the simple-band model. Without
the central-cell corrections, which here account for the deviations
of the exciton ground state from the hydrogen-like series, the solutions
of $H_{\mathrm{sb}}$ are given by
\begin{equation}
E_{n}=E_{\mathrm{g}}-\frac{R_{\mathrm{exc}}}{n^{2}}
\end{equation}
with the principal quantum number $n$ and the excitonic Rydberg energy
$R_{\mathrm{exc}}$~\cite{SO}. The eigenfunctions are the well-known
solutions of the Schr\"odinger equation of the hydrogen atom, where
only the Bohr radius $a_{0}$ is to be replaced by the excitonic radius
$a_{\mathrm{exc}}=\varepsilon\gamma'_{1}a_{0}$~\cite{SO}. 

The spin-orbit interaction $H_{\mathrm{so}}$ couples the quasi-spin
$I=1$ and the hole spin $S_{\mathrm{h}}=1/2$ to the effective
hole spin $J=I+S_{\mathrm{h}}$, where $J=1/2$ corresponds
to the $\Gamma_{7}^{+}$ valence band and $J=3/2$ corresponds
to the $\Gamma_{8}^{+}$ valence bands. The value of $J$ therefore
distinguishes between the yellow $(J=1/2)$ and
the green $(J=3/2)$ exciton series (Fig.~\ref{fig:Band-structure-of}).
Within this approximation, these are two noninteracting hydrogen-like
exciton series.

The remaining parts of $H$ without the exchange interaction form
the so-called $H_{d}$~term:
\begin{widetext}
\begin{equation}
H_{d}=\frac{\gamma'_{1}}{2\hbar^{2}m_{0}}\left\{ -\frac{\mu'}{3}P^{(2)}\cdot I^{(2)}+\frac{\delta'}{3}\left(\sum_{k=\pm4}\left[P^{(2)}\times I^{(2)}\right]_{k}^{(4)}+\frac{\sqrt{70}}{5}\left[P^{(2)}\times I^{(2)}\right]_{0}^{(4)}\right)\right\}.
\end{equation}
\end{widetext}
This term mixes the two exciton series as discussed in Sec.~\ref{sub:VBS}. 
In the spherical approximation
$(\delta'=0)$, in which the Hamiltonian has still spherical symmetry,
the momentum $F=L+J$ and its $z$-component
$M_{F}$ are good quantum numbers, while $\boldsymbol{L}$ and $\boldsymbol{J}$
do not commute with $H_{d}$. This leads to a fine-structure splitting
of the eigenstates of the Hamiltonian, which is discussed, e.g., in
Refs.~\cite{7,7_11,17_17} for several semiconductors. The angular
momentum part of an appropriate basis set reads
\begin{equation}
\left|L;\,\left(I,\, S_{\mathrm{h}}\right),\, J;\, F,\, M_{F}\right\rangle \left|S_{\mathrm{e}},\, M_{S_{\mathrm{e}}}\right\rangle ,\label{eq:basis1}
\end{equation}
where the $z$-component $M_{S_{\mathrm{e}}}$ of the electron spin
$S_{\mathrm{e}}=1/2$ is also a good quantum number.

The other parts of $H_{d}$ with the coefficient $\delta'$ have
cubic symmetry. For this reason neither $F$ nor $M_{F}$ are good
quantum numbers anymore \cite{17_17_26}. The eigenstates of the
Hamiltonian transform according to the irreducible representation
$\Gamma_{i}^{\pm}$ of the cubic group $O_{\mathrm{h}}$ instead of
those of the full rotation group. 

In the case in which the value of an arbitrary (integral or half-integral
valued) momentum $A$ is less than or equal to four, it is possible
to form linear combinations of the states $\left|A,\, M_{A}\right\rangle $
that transform according to the irreducible representations of $O_{\mathrm{h}}$~\cite{G1,28}.
For example, the state $\left(\left|3,\,2\right\rangle -\left|3,\,-2\right\rangle \right)/\sqrt{2}$
transforms according to the irreducible representation $\Gamma_{2}^{-}$
of $O_{\mathrm{h}}$. These states are often denoted by $\left|A,\,\Gamma_{i}\right\rangle $~\cite{17_17_26}.
However, this procedure is not uniquely possible for $A>4$ due to
arising degeneracies~\cite{G1}. Therefore, it is reasonable to describe
the angular-momentum part by Eq.~(\ref{eq:basis1}) even if $\delta'\neq0$
holds. 

The effect of $H_{d}$ on the eigenstates of the Hamiltonian
decreases with increasing principal quantum number $n$ since the
wave functions extend over more unit cells and the cubic symmetry
of the solid becomes less important~\cite{28}. An approach to treat
the effects of $H_{d}$ on exciton states with high values of 
$n$ in a simple way can be found in Ref.~\cite{80}.

The exchange interaction $H_{\mathrm{exch}}$ couples the spins of electron
and hole and leads to a splitting of $S$ excitons, i.e., excitons
with $L=0$, into ortho and para excitons~\cite{7}. The coupling of
the spins leads to a total momentum $F_{t}=F+S_{\mathrm{e}}$ and
we finally obtain
\begin{equation}
\left|L;\,\left(I,\, S_{\mathrm{h}}\right),\, J;\, F,\, S_{\mathrm{e}};\, F_{t},\, M_{F_{t}}\right\rangle 
\end{equation}
for the angular momentum part of an appropriate basis set.

In the literature the radial part of the basis is often not specified.
A typical ansatz for the wavefunction of the exciton is
\begin{equation}
\Psi=\sum_{\beta}g_{\beta}\left(r\right)\left|L;\,\left(I,\, S_{\mathrm{h}}\right),\, J;\, F,\, S_{\mathrm{e}};\, F_{t},\, M_{F_{t}}\right\rangle ,
\end{equation}
where $\beta$ denotes the quantum numbers $L$, $J$, $F$, $F_{t}$,
and $M_{F_{t}}$. The radial functions $g_{\beta}\left(r\right)$
are often determined using finite-element methods~\cite{7,7_13} or
variational methods~\cite{7_11,17_17,17,17_17_26,17_15}. Unfortunately,
these methods lead to a huge number of coupled differential equations
for the functions $g_{\beta}\left(r\right)$.

In contrast to earlier works, in this paper we use a complete basis for
the radial functions. Since the eigenfunctions of the hydrogen atom
do not represent a complete basis without the continuum states, we
use the so-called Coulomb-Sturmian functions as described, e.g., in
Ref.~\cite{S1}. The radial functions of this basis read
\begin{equation}
U_{NL}\left(r\right)=N_{NL}\left(2\rho\right)^{L}e^{-\rho}L_{N}^{2L+1}\left(2\rho\right)\label{eq:U}
\end{equation}
with $\rho=r/\alpha$, a normalization factor $N_{NL}$,
the associated Laguerre polynomials $L_{n}^{m}\left(x\right)$, and
an arbitrary scaling parameter $\alpha$. Note that we here use the
radial quantum number $N$, which is related to the principal quantum
number $n$ via $n=N+L+1$. 
Various recursion relations of these functions, which are needed for
our calculations, are given in Appendix~\ref{sub:Recursion-relations-of}.

Our basis set finally reads
\begin{eqnarray}
\left|\Pi\right\rangle  & = & \left|N,\, L;\,\left(I,\, S_{\mathrm{h}}\right),\, J;\, F,\, S_{\mathrm{e}};\, F_{t},\, M_{F_{t}}\right\rangle \label{eq:basis}
\end{eqnarray}
and we make the ansatz
\begin{equation}
\left|\Psi\right\rangle =\sum_{NLJFF_{t}M_{F_{t}}}c_{NLJFF_{t}M_{F_{t}}}\left|\Pi\right\rangle \label{eq:ansatz}
\end{equation}
with real coefficients $c$. Since the functions $U_{NL}\left(r\right)$
actually depend on the coordinate $\rho=r/\alpha$, we substitute
$r\rightarrow\rho\alpha$ in the Hamiltonian (\ref{eq:H0}) and multiply
the corresponding Schr\"odinger equation $H\Psi=E\Psi$ by $\alpha^{2}$.
Then we calculate a matrix representation of the Schr\"odinger equation,
which yields a generalized eigenvalue problem of the form
\begin{equation}
\boldsymbol{A}\boldsymbol{c}=E\boldsymbol{M}\boldsymbol{c},\label{eq:gev}
\end{equation}
which is solved in atomic units using an appropriate LAPACK routine~\cite{Lapack}.
The matrix elements which enter the symmetric matrices $\boldsymbol{A}$
and $\boldsymbol{M}$ are given in Appendices~\ref{sub:Matrix-elements}
and~\ref{sub:Reduced-matrix-elements}. The vector $\boldsymbol{c}$
contains the coefficients of the ansatz (\ref{eq:ansatz}). Since
the basis cannot be infinitely large, the values of the quantum numbers
are chosen in the following way: For each value of $n=N+L+1$ we use
\begin{eqnarray}
L & = & 0,\,\ldots,\, n-1,\nonumber \\
J & = & 1/2,\,3/2,\nonumber \\
F & = & \left|L-J\right|,\,\ldots,\,\min\left(L+J,\, F_{\mathrm{max}}\right),\\
F_{t} & = & F-1/2,\, F+1/2,\nonumber \\
M_{F_{t}} & = & -F_{t},\,\ldots,\, F_{t}.\nonumber 
\end{eqnarray}

The value $F_{\mathrm{max}}$ and the maximum value of $n$ are chosen
appropriately large so that the eigenvalues converge. Additionally,
we can use the scaling parameter $\alpha$ to enhance convergence.
In particular, if the eigenvalues of excitonic states with principal
quantum number $n$ are calculated, we set 
$\alpha=na_{\mathrm{exc}}$ according to Ref.~\cite{S1}.

\subsection{Oscillator strengths \label{sub:Oscillator-strengths-1}}

If no external fields are present, the different terms of the Hamiltonian
couple only basis states with even or with odd values of $L$ (see
Appendix~\ref{sub:Matrix-elements}). Since we restrict ourselves
to odd values of $L$, the exchange interaction and central-cell corrections
can be neglected. 

$F$ as well as $M_{F}$ are good quantum numbers
in the spherical approximation. In this case the usage of the total
momentum $F_{t}$ in our basis (\ref{eq:basis}) seems unnecessary.
However, we still keep the total momentum since it allows us to determine
the symmetry of the different excitonic states.

\textcolor{black}{If the spins of the electron and the hole are considered in the simple-band model,
the exciton states are either spin-singlet or spin-triplet states.
Since the spin is conserved in dipole transitions, the oscillator strength is nonzero only for the singlet states.
However, an optical excitation at the $\Gamma$-point $(\boldsymbol{k}=\boldsymbol{0})$ is forbidden in $\mathrm{Cu_{2}O}$
since the valence band and the conduction band have the same parity.
Only $\boldsymbol{k}$-dependent admixtures to the transition matrix element enable an optical excitation. The leading term in 
this matrix element is therefore proportional
to the $\boldsymbol{k}$-vector in Fourier space or to the gradient 
(of the envelope function) 
in position space at $r=0$.
As the $\boldsymbol{k}$-vector transforms according to the irreducible representation $D^{1}$ of
the rotation group, the envelope function of the exciton has to transform according to the same representation.
Since $L$ is a good quantum number in the simple-band model, 
only $P$-excitons, i.e., excitons with $L=1$, are dipole-allowed in this case~\cite{GRE}.}

\textcolor{black}{The oscillator strength is then nonzero
only if the total symmetry of the exciton,
which is given by the symmetry of envelope function
and the symmetries of the bands,
\begin{equation}
\Gamma_{\mathrm{exc}}=\Gamma_{\mathrm{env}}\otimes\Gamma_{\mathrm{c}}\otimes\Gamma_{\mathrm{v}},
\end{equation}
is identical to the symmetry $\Gamma_{4}^{-}$ of the dipole operator~\cite{G1,TOE}.}

Since we consider the symmetry of the valence band via the spins $I$
and $S_{\mathrm{h}}$ as well as the symmetry of the conduction band
via the spin $S_{\mathrm{e}}$ in our basis, the total symmetry of
the exciton can immediately be obtained by an examination of the solutions
of the Schr\"odinger equation in the complete basis set of Eq.~(\ref{eq:basis}).
Three important points have to be considered in this examination:

We already stated in Sec.~\ref{sub:Basis} that it is possible
to combine the different states $\left|A,\, M_{A}\right\rangle $
of an arbitrary momentum $A\leq4$ to the states $\left|A,\,\Gamma_{i}\right\rangle $.
Solving the eigenvalue problem (\ref{eq:gev}), we obtain the coefficients
$c$ of the basis functions according to Eq.~(\ref{eq:ansatz}).
We can now compare the coefficients of those basis functions with
a fixed value of $F_{t}\leq4$ to the coefficients of the functions
$\left|A,\,\Gamma_{i}\right\rangle $ given in Ref.~\cite{G1} to obtain
the symmetry of the eigenstates.

If an eigenvalue is $p$-fold degenerate, one has to form appropriate
linear combinations of the $p$ eigenvectors $\boldsymbol{c}$ at
first, before a comparison with the coefficients of 
the states $\left|A,\,\Gamma_{i}\right\rangle $
is possible.

The quasi-spin $I$ transforms according to $\Gamma_{5}^{+}$ whereas
a normal spin one transforms according to $\Gamma_{4}^{+}$. Since
$\Gamma_{5}^{+}=\Gamma_{2}^{+}\otimes\Gamma_{4}^{+}$ holds for the
cubic group $O_{\mathrm{h}}$ \cite{G1}, one has to multiply the symmetries
$\Gamma_{i}$ obtained via the above comparison by $\Gamma_{2}^{+}$~\cite{28}.

For the states of $\Gamma_{4}^{-}$-symmetry we can then calculate
relative oscillator strengths:
With the above explanations, the dipole-allowed states must have a nonvanishing overlap with the state~\cite{G1}
\begin{eqnarray}
\left|D\right\rangle & = &\frac{1}{\sqrt{2}}\left(\left|\left(\frac{1}{2},\,\frac{1}{2}\right)\,0,\,1;\,1,\,1;\,2,\,2\right\rangle\right.\nonumber \\
 & & \nonumber \\
 & & \qquad\quad\left.-\left|\left(\frac{1}{2},\,\frac{1}{2}\right)\,0,\,1;\,1,\,1;\,2,\,-2\right\rangle\right),
\end{eqnarray}
where the quantum numbers denote the angular momenta in $\left|\left(S_{\mathrm{e}},\,S_{\mathrm{h}}\right)\,S,\,I;\,I+S,\,L;\,F_t,\,M_{F_t}\right\rangle$.
This state transforms also according to the irreducible representation $\Gamma_{4}^{-}$. 
Its specific form shows that we assume the incident light to be linearly polarized~\cite{G1}.
The relative oscillator strengths are finally given by
\begin{equation}
f_{\mathrm{rel}}\sim\left|\lim_{r\rightarrow0}\frac{\partial}{\partial r}\left\langle D\middle|\Psi\left(\boldsymbol{r}\right)\right\rangle\right|^2
\end{equation}
with the wave function of Eq.~(\ref{eq:ansatz}) in spatial representation
(see Appendix~\ref{sub:Oscillator-strengths}).

\section{Results and discussion}

\subsection{$F$=5/2 and 7/2 excitonic lines of cuprous oxide \label{sub:Determiantion-of}}

In this section we apply the method presented in Sec.~\ref{sec:Theory} and repeat the 
analysis of Ref.~\cite{28}.
Deviations from the results in Ref.~\cite{28} are observed and discussed.
Using the parameters $\Delta=0.1338\,\mathrm{eV}$,
$\gamma_{1}'=2.78$, $\mu'=0.47$ from Ref.~\cite{7},
the parameter $\delta'$ has been determined in Ref.~\cite{28} by
comparing theoretical results with experimental absorption spectra.

In the spherical approximation $F$ and
$M_{F}$ are good quantum numbers. Including the cubic part of the
Hamiltonian $H_{d}$, the reduction of the irreducible representations
$D^{F}$ of the rotation group by the cubic group $O_{\mathrm{h}}$
has to be considered~\cite{G1}. With the additional factor $\Gamma_{2}^{+}$
as described in Sec.~\ref{sub:Oscillator-strengths-1}, we obtain
for the symmetry of the envelope function and the hole
\begin{subequations}
\begin{align}
\tilde{D}^{\frac{1}{2}} = &\: D^{\frac{1}{2}}\otimes\Gamma_{2}^{+}=\Gamma_{6}^{-}\otimes\Gamma_{2}^{+}=\Gamma_{7}^{-},\label{eq:12}\\
\displaybreak[1]
\nonumber \\
\tilde{D}^{\frac{3}{2}} = &\: D^{\frac{3}{2}}\otimes\Gamma_{2}^{+}=\Gamma_{8}^{-}\otimes\Gamma_{2}^{+}=\Gamma_{8}^{-},\\
\displaybreak[1]
\nonumber \\
\tilde{D}^{\frac{5}{2}} = &\: D^{\frac{5}{2}}\otimes\Gamma_{2}^{+}=\left(\Gamma_{7}^{-}\oplus\Gamma_{8}^{-}\right)\otimes\Gamma_{2}^{+}\nonumber \\
&\: \qquad\qquad\,\,=\Gamma_{6}^{-}\oplus\Gamma_{8}^{-},\label{eq:52}\\
\displaybreak[1]
\nonumber \\
\tilde{D}^{\frac{7}{2}} = &\: D^{\frac{7}{2}}\otimes\Gamma_{2}^{+}=\left(\Gamma_{6}^{-}\oplus\Gamma_{7}^{-}\oplus\Gamma_{8}^{-}\right)\otimes\Gamma_{2}^{+}\nonumber \\
&\: \qquad\qquad\,\,=\Gamma_{7}^{-}\oplus\Gamma_{6}^{-}\oplus\Gamma_{8}^{-}.\label{eq:72}
\end{align}
\end{subequations}
The Hamiltonian couples only states with even or odd values of
$L$. Since in the simple-band model only states with $L=1$ are dipole-allowed,
we only include states with odd values of $L$ in our basis. This
is the reason for the negative parities in Eqs.~(\ref{eq:12})-(\ref{eq:72}).
Furthermore, we can neglect the central-cell corrections and the exchange 
interaction in the following since they only affect states with $L=0$.

\begin{figure}
\begin{centering}
\includegraphics[width=1.0\columnwidth]{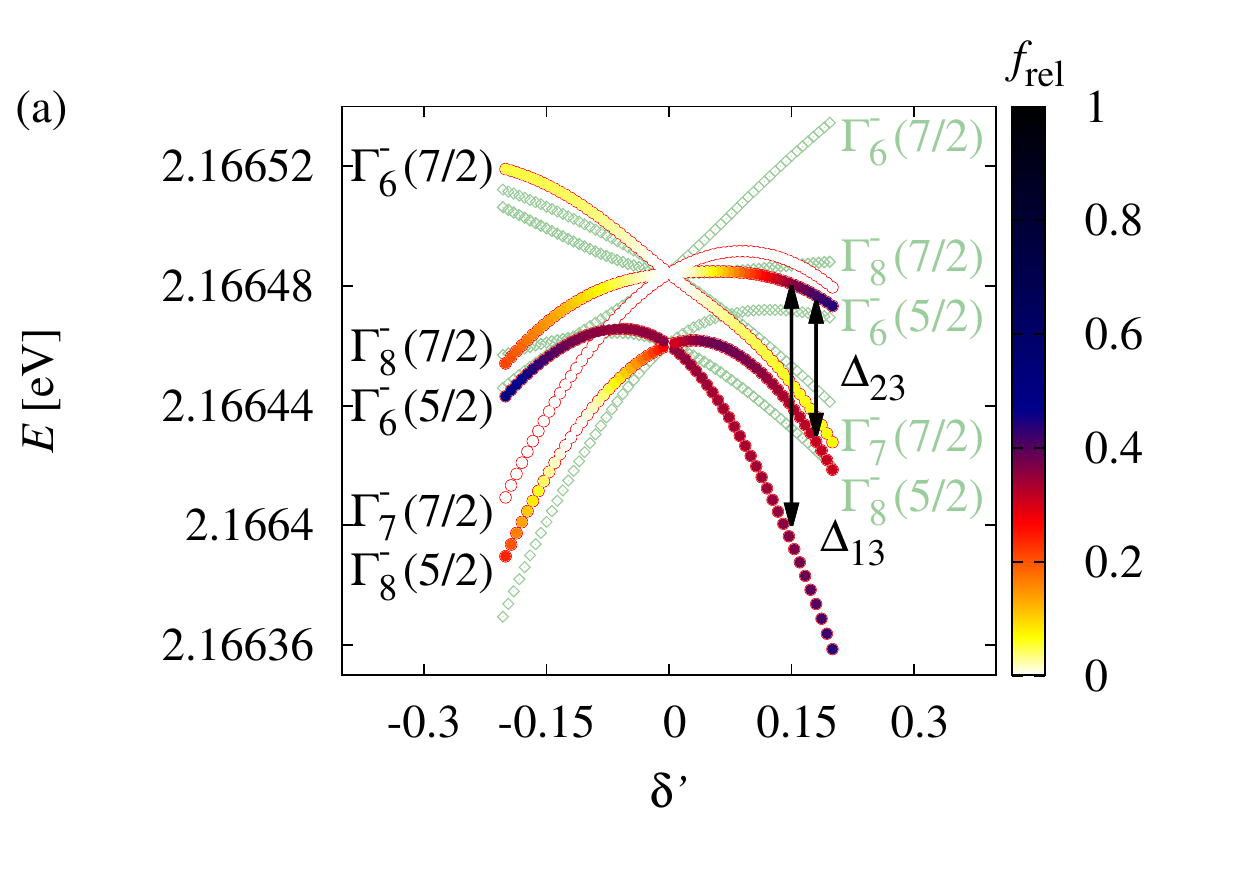}

\includegraphics[width=0.9\columnwidth]{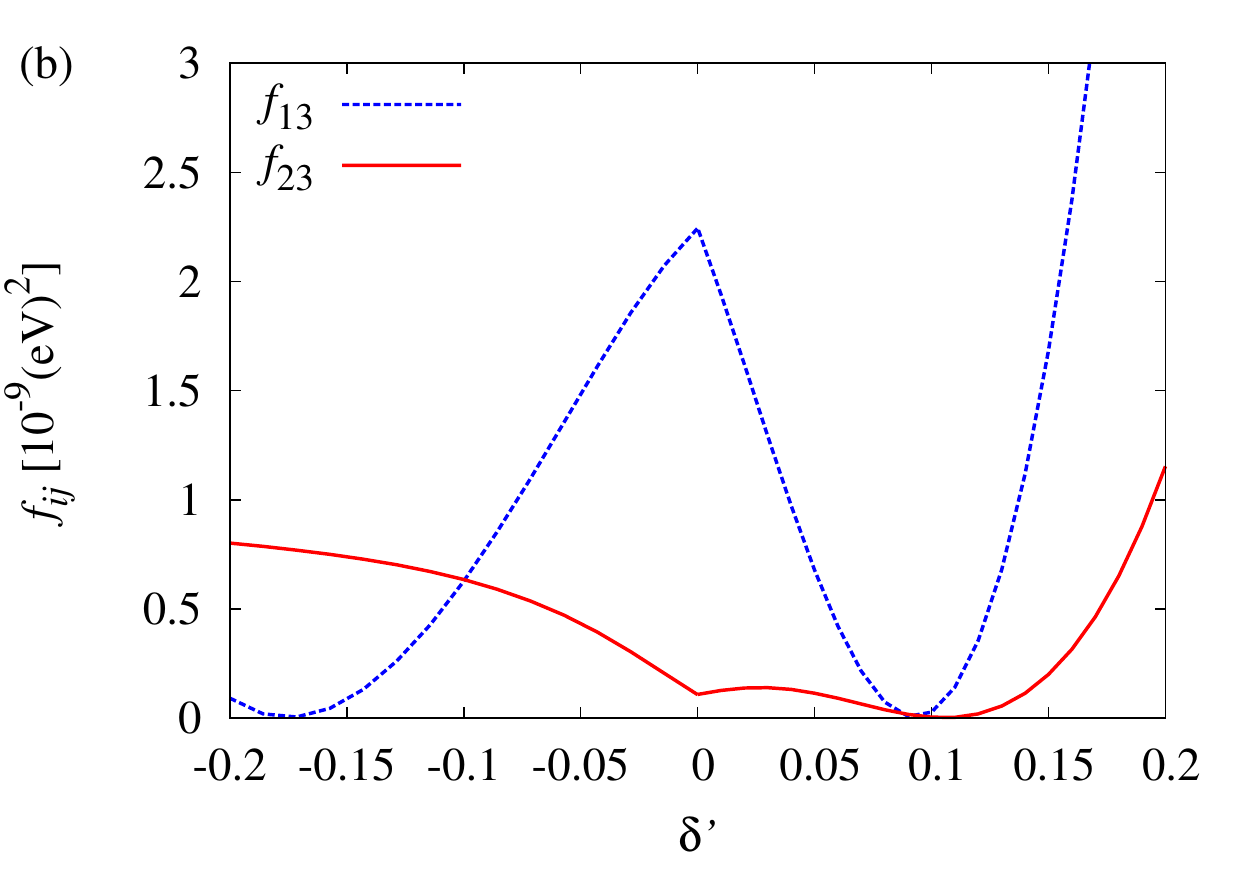}
\par\end{centering}

\protect\caption{(Color online) (a) Comparison of the results of Ref.~\cite{28} 
(green diamonds) and the results obtained
by the method described in Sec.~\ref{sec:Theory} (circles). The splitting
of the $F=5/2$ and $F=7/2$ states due to the cubic
part of the Hamiltonians is depicted for the principal quantum number
$n=4$. Our symmetry assignment (left) differs from the one of Ref.~\cite{28}
(right). The color bar shows the relative oscillator strengths 
in arbitrary units. The state assigned with $\Gamma_{6}^{-}\left(7/2\right)$
has only small oscillator strength. 
The parameters $\Delta_{13}$ and $\Delta_{23}$ denote the line spacings
between the remaining dipole-allowed states. 
(b) The functions $f_{13}(\delta')$ and $f_{23}(\delta')$
as defined in Eq.~(\ref{eq:fij}). The value $\delta'\approx0.1$,
for which both function values are minimal, is the value of this
material parameter in $\mathrm{Cu_{2}O}$ if $\mu'=0.47$ and $\eta_i=0$ holds. 
The kinks at $\delta'=0$
are due to the definition of the quantities $\Delta_{13}$ 
and $\Delta_{23}$.
For further information see text. \label{fig:Oscillator-strengths-of}}
\end{figure}

As can be seen from Eqs.~(\ref{eq:52}) and (\ref{eq:72}) the states
with $F=5/2$ and $F=7/2$ split into five states
with the symmetries $\Gamma_{6}^{-}$, $\Gamma_{7}^{-}$, and $\Gamma_{8}^{-}$.
The degeneracies of these states are two for $\Gamma_{6}^{-}$ and
$\Gamma_{7}^{-}$ and four for $\Gamma_{8}^{-}$. Including the symmetry
$\Gamma_{6}^{+}$ of the electron spin, we obtain the symmetry of
the exciton and can determine which of these states are dipole-allowed, viz.,
\begin{subequations}
\begin{align}
\Gamma_{6}^{-}\otimes\Gamma_{6}^{+} = &\: \Gamma_{1}^{-}\oplus\Gamma_{4}^{-},\\
\displaybreak[1]
\nonumber \\
\Gamma_{7}^{-}\otimes\Gamma_{6}^{+} = &\: \Gamma_{2}^{-}\oplus\Gamma_{5}^{-},\\
\displaybreak[1]
\nonumber \\
\Gamma_{8}^{-}\otimes\Gamma_{6}^{+} = &\: \Gamma_{3}^{-}\oplus\Gamma_{4}^{-}\oplus\Gamma_{5}^{-}.
\end{align}
\end{subequations}
Since only the threefold degenerate states of symmetry $\Gamma_{4}^{-}$
are dipole-allowed, four lines are visible in experiments at the most.
However, in Ref.~\cite{28} only three lines could be observed.

We solve the generalized eigenvalue problem (\ref{eq:gev})
in the complete basis of Eq.~(\ref{eq:basis}) 
with the additional quantum number $F_{t}$, determine the states of symmetry
$\Gamma_{4}^{-}$, and calculate the oscillator strengths. Even though
the states with $F=5/2$ and $F=7/2$ lie energetically
very close together in the spherical approximation, it is inappropriate
to consider only basis functions with these two values of $F$ in our ansatz.
Including all states with $F\leq15/2$, we obtain a clearly
different result in comparison to the one of Ref.~\cite{28}. 
In Fig.~\ref{fig:Oscillator-strengths-of}(a) we depict the results of
Ref.~\cite{28} for
the principal quantum number $n=4$
by green diamonds and our results by circles.
For the $9j$-symbol in Eq.~(\ref{eq:9j13}) we use the relations given 
in Refs.~\cite{ED,86}, so that our result differs by an exchange of 
the first two rows in the $9j$-symbol of Eq.~(14) of Ref.~\cite{28}
\textcolor{black}{or of Eq.~(A2) of Ref.~\cite{17_17_26}. We 
are convinced that the formulas given in Refs.~\cite{ED,86} are correct as the 
quantum numbers in the rows of the $9j$-symbol appear in the same order as they appear in
our basis states.
An odd permutation of rows can lead to a change in the sign of the 
$9j$-symbol depending on the quantum numbers included~\cite{ED} (cf. also Appendix~\ref{sub:Matrix-elements}).}
Therefore, our
assignment of the lines with the symmetries 
$\Gamma_{6}^{-}$, $\Gamma_{7}^{-}$, $\Gamma_{8}^{-}$
in Fig.~\ref{fig:Oscillator-strengths-of}(a) differs from the one of Ref.~\cite{28}.
The oscillator strengths of the states calculated are also depicted in
this figure. 

In Ref.~\cite{28} it has
been discussed that also components of the order $\boldsymbol{p}^{4}$
should be included in the Hamiltonian to obtain more reliable values
for the oscillator strengths. However, the effect of these terms is
considered to be very weak for the states investigated here and is
generally only important for the $1S$ exciton state~\cite{1}.
Indeed, the effect of $\boldsymbol{p}^{4}$-terms significantly decreases
with increasing principal quantum number $n$ but a corresponding decrease
of the oscillator strength between the $n=4$ states and the $n=5$
states cannot be observed experimentally~\cite{28}. This shows that
an effect of $\boldsymbol{p}^{4}$-terms is not present at all. We
therefore neglect these higher order terms in $\boldsymbol{p}$.

In Fig.~\ref{fig:Oscillator-strengths-of}(a)
the state assigned with $\Gamma_{6}^{-}\left(7/2\right)$
has only small oscillator strength, which validates the fact that
only three lines can be observed experimentally in absorption spectra.
Furthermore, the two lines assigned with $\Gamma_{6}^{-}\left(7/2\right)$ and
$\Gamma_{8}^{-}\left(5/2\right)$ could hardly be resolved in 
ex-periments for $\delta'\geq 0.1$.

\begin{figure}
\begin{centering}
\includegraphics[width=1.0\columnwidth]{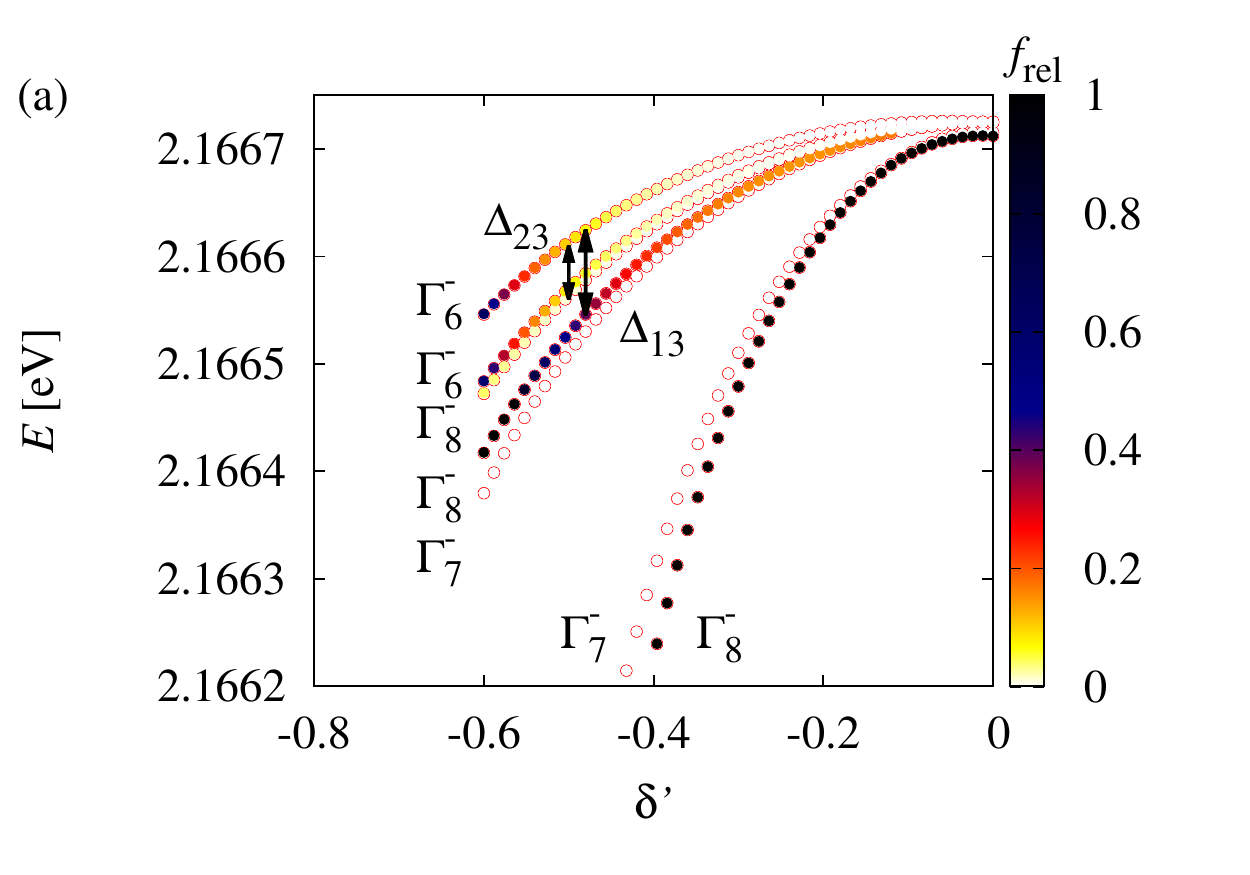}

\includegraphics[width=0.9\columnwidth]{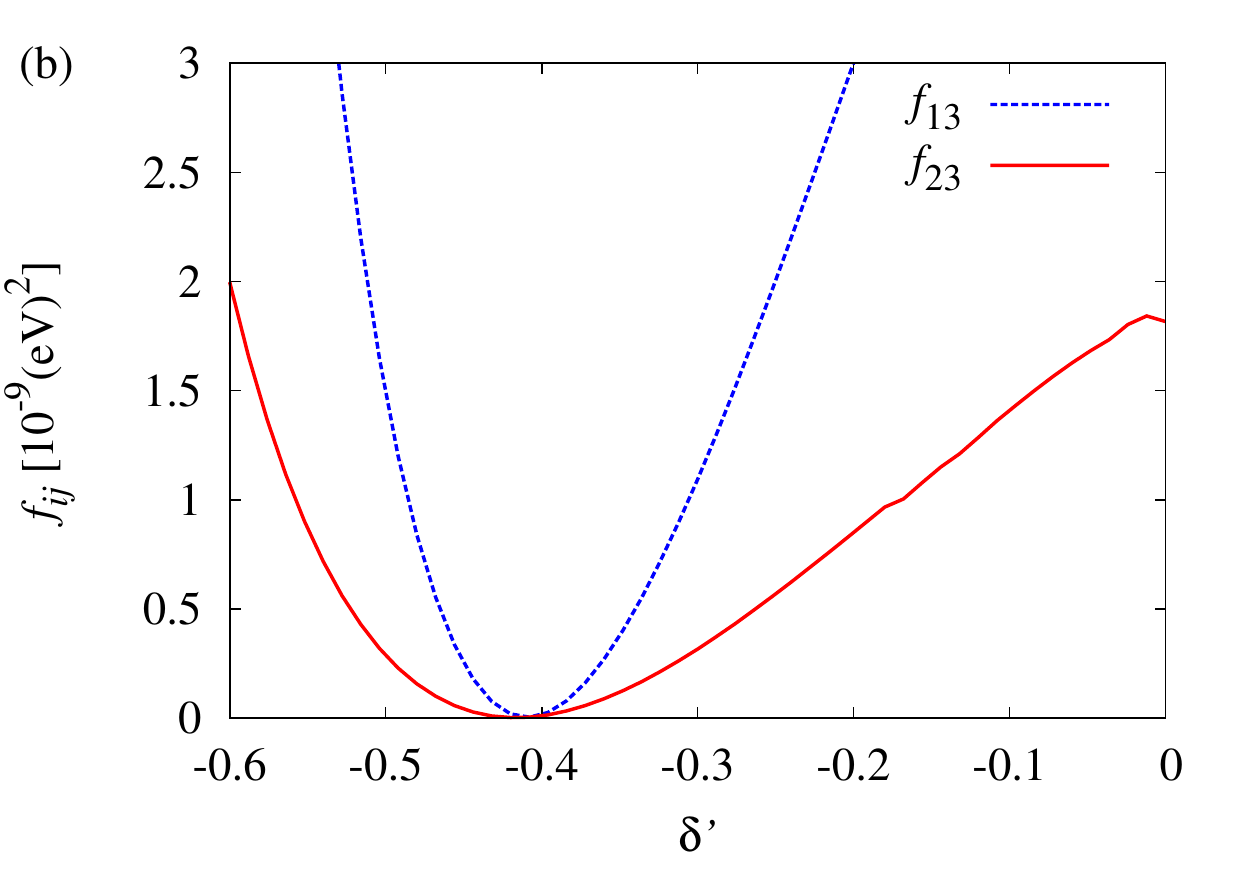}
\par\end{centering}

\protect\caption{(Color online) Same calculation as in Fig.~\ref{fig:Oscillator-strengths-of} but
for the material parameters $\Delta$, $\gamma_{1}'$,
$\mu'$, and $eta_i$ obtained from band structure calculations~\cite{20,80}
(cf. Table~\ref{tab:1}).
(a) Relative oscillator strengths
(color bar) of the exciton states with
$n=4$ in arbitrary units. The assignment of $F$ quantum numbers is omitted (see text).
(b) The functions $f_{13}(\delta')$ and $f_{23}(\delta')$
as defined in Eq.~(\ref{eq:fij}). 
The optimum value for $\delta'$ is here $\delta'=-0.408$.
For further information see text.~\label{fig:Oscillator-strengths-of1}}
\end{figure}

In order to determine the correct value of $\delta'$ we consider
the energetic spacing between the lines. We use the same nomenclature
as in Ref.~\cite{28}; i.e., the spacing between the state with the
highest energy and the state with the lowest energy is called $\Delta_{13}$
while the spacing between the state with the highest energy and the
state with the second highest energy is called $\Delta_{23}$ [see
Fig.~\ref{fig:Oscillator-strengths-of}(a)]. 
Note that, e.g., the state with the lowest energy is $\Gamma_{8}^{-}\left(5/2\right)$ 
for $\delta'<0$ and $\Gamma_{6}^{-}\left(5/2\right)$ for $\delta'>0$ so 
that there are different lines entering $\Delta_{13}$ and $\Delta_{23}$
depending on $\delta'$.
Since the spacings
depend on the value of $\delta'$ and the principal quantum number
$n$, we use the notation $\Delta_{ij}(\delta',\, n)$.
Minimizing the functions
\begin{equation}
f_{ij}(\delta')=\sum_{n=4}^{8}\left(\Delta_{ij}(\delta',\, n)-\Delta_{ij}^{\mathrm{exp}}(n)\right)^{2},\label{eq:fij}
\end{equation}
where $\Delta_{ij}^{\mathrm{exp}}(n)$ denote the spacings
in the experimental absorption spectrum, we obtain almost the same
value of $\delta'$ irrespective of the indices $ij$ as can be
seen from Fig.~\ref{fig:Oscillator-strengths-of}(b). The final value is
\begin{equation}
\delta'=0.1,
\end{equation}
which is clearly different from the value $\delta'=-0.1$ of Ref.~\cite{28}.

\begin{figure}
\begin{centering}
\includegraphics[width=0.9\columnwidth]{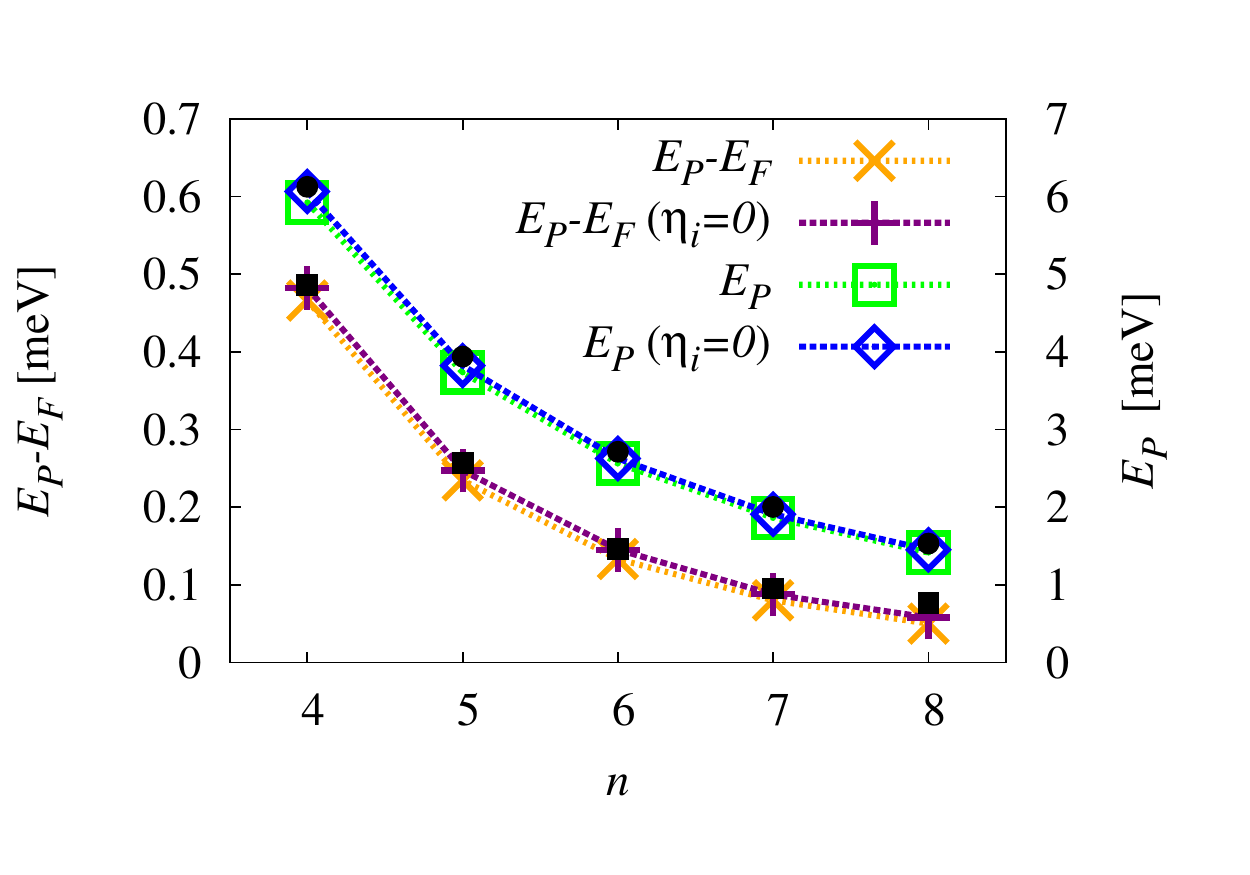}
\par\end{centering}

\protect\caption{(Color online) Binding energies $E_P$ of the dipole-allowed 
$P$ excitons and the energy difference between these 
excitons and the average energy of the dipole-allowed $F$ excitons.
Black dots and circles denote experimental data.
Theoretical results are depicted by linespoints.
The best agreement between theory and experiment is obtained for $\delta'\approx-0.42$.
Without the $\eta_i$-dependent terms of Eq.~(\ref{eq:Hh})
an even better agreement can be obtained for $\delta'=-0.408$.
\label{fig:PF}}
\end{figure}

Even though the values $\mu'=0.47$ and $\delta'=0.1$ reproduce the 
experimental results of excitonic absorption spectra very well,
we cannot disregard that these values originate from the 
valence band structure of $\mathrm{Cu_{2}O}$.
In Ref.~\cite{28} it has already been noted that a negative value
of $\delta'$ is expected due to a comparison with band structure 
calculations~\cite{20}.
However, our calculations do not provide a negative value 
even if we include the state assigned
with $\Gamma_{6}^{-}\left(7/2\right)$ in our calculations.

A fit to band structure 
calculations yields~\cite{80} 
$\mu'=0.0586$ and $\delta'=-0.404$ as already mentioned 
in Sec.~\ref{sub:Hamiltonian}. These values are clearly different
from the results of the above calculation.
Therefore, we assume that there is more than one 
combination of the parameters $\mu'$ and $\delta'$ 
yielding the correct spacings between the observed exciton states.

To prove our assumption, we perform the same analysis again, using the
parameters $\Delta=0.131\,\mathrm{eV}$, $\gamma_{1}'=2.77$, and
$\mu'=0.0586$ as given values. We now also include the terms with the coefficients 
$\eta_i$ of the Hamiltonian~(\ref{eq:Hh}) of the hole.
We restrict the analysis to negative values of $\delta'$.
The results are depicted in Fig.~\ref{fig:Oscillator-strengths-of1}.
Note that the value of $\mu'=0.0586$ is so small that the 
states with different quantum number $F$ are hardly separated for $\delta'=0$.
Since these states mix for finite $\delta'$ it is therefore inappropriate to give 
the symmetries of the states in the form $\Gamma^{-}_{i}(F)$ and we omit the
assignment with $F$.
Furthermore, the spacing between the lower lying $\Gamma^{-}_{6}$ state and 
the higher lying $\Gamma^{-}_{8}$ state, which has a very small oscillator strength, is so
small that it can hardly be resolved in experiments.
This proves again that there are only three lines observable in experiments.
Our calculations yield
\begin{equation}
\delta'=-0.408,
\end{equation}
which is in excellent agreement with the expected value of $\delta'=-0.404$.

In Fig.~\ref{fig:PF} we depict the binding energies $E_P$ of the dipole-allowed $P$ excitons
as well as the energy difference between these excitons 
and the average energy of the dipole-allowed $F$ excitons.
We use the nomenclature of Ref.~\cite{28} so that the
$P$ excitons are the energetically 
lower lying states of symmetry $\Gamma^{-}_7$ and $\Gamma^{-}_8$ in
Fig.~\ref{fig:Oscillator-strengths-of1}(a) whereas the $F$ excitons
are the remaining states in this Figure.
We obtain a good agreement between theory and experiment for a 
slightly different value of $\delta'\approx-0.42$.

Small uncertainties in the material parameters still remain but can be 
explained in terms of some approximations made, e.g., 
the parameters $A_i$ and $B_i$ taken from Ref.~\cite{80}
are fit parameters to band structure calculations and are hence afflicted with errors.
Another influence on the exciton spectrum are phonons~\cite{75}, 
which our theory does not account for and which also make 
an experimental determination of the correct position 
of exciton resonances difficult~\cite{GRE}.
Finally, we think that out of the several combinations of $\mu'$ and $\delta'$,
which reproduce the exciton spectrum,
the parameters obtained from band structure calculations are the 
correct ones to describe this spectrum.

\subsection{$K$-dependent line splitting \label{sub:-dependent-line-splitting}}

In this section we discuss the $K$-dependent line splitting observed in Ref.~\cite{9_1}
in terms of the complex valence band structure of $\mathrm{Cu_{2}O}$.
The Hamiltonian~(\ref{eq:Hpeph}) depends
only on the relative coordinate $\boldsymbol{r}=\boldsymbol{r}_{\mathrm{e}}-\boldsymbol{r}_{\mathrm{h}}$
of electron and hole. For this reason the momentum of the center of
mass $\hbar\boldsymbol{K}$ is a constant of motion~\cite{24}. About
ten years ago the $K$-dependent line splitting of the $1S$ exciton
state was observed and explained in terms of a $K$-dependent
short-range exchange interaction of the form~\cite{9_1,8,9} 
\begin{widetext}
\begin{equation}
J\left(\boldsymbol{K}\right)=\Delta_{1}\left(\begin{array}{ccc}
K^{2} & 0 & 0\\
0 & K^{2} & 0\\
0 & 0 & K^{2}
\end{array}\right)+\Delta_{3}\left(\begin{array}{ccc}
3K_{1}^{2}-K^{2} & 0 & 0\\
0 & 3K_{2}^{2}-K^{2} & 0\\
0 & 0 & 3K_{3}^{2}-K^{2}
\end{array}\right)+\Delta_{5}\left(\begin{array}{ccc}
0 & K_{1}K_{2} & K_{1}K_{3}\\
K_{1}K_{2} & 0 & K_{2}K_{3}\\
K_{1}K_{3} & K_{2}K_{3} & 0
\end{array}\right),
\label{eq:excans}
\end{equation}
\end{widetext}
Fitting this ansatz to experimental spectra of the $1S$ exciton yielded
\begin{equation}
\Delta_1=-8.6~\mathrm{\mu eV},\quad\Delta_3=-1.3~\mathrm{\mu eV},\quad\Delta_5=2~\mathrm{\mu eV}.
\end{equation}
However, it has been reported that a
$K$-dependent short-range exchange interaction is far too small to cause the
large line splitting observed~\cite{1}.

As has already been stated in Sec.~\ref{sub:VBS},
when considering the ortho exciton states
each of these states can have a 
larger contribution of one of the orbital Bloch functions $xy$, $yz$, or $zx$
without a violation of symmetry, respectively.
However, if one orbital Bloch component predominates, the anisotropic 
dispersion of the Bloch function will lead to an
anisotropic dispersion of the excitons. Thus, the $K$-dependent 
line splitting of the ortho exciton observed in 
Refs.~\cite{9_1,8,9} should be a direct consequence of the disparity 
in the orbital Bloch components of this exciton.
Therefore, we think that the $I$-dependent terms in Eq.~(\ref{eq:Hpeph}) 
are the reason for this splitting and we will show that these terms are 
of the same form as the ones in Eq.~(\ref{eq:excans}). Since $\eta_i\ll\gamma_i$ holds
in Eq.~(\ref{eq:Hh}), we set $\eta_i=0$ in the following.

Inserting the well-known coordinates and momenta 
of relative and center of mass motion,
\begin{subequations}
\begin{align}
\boldsymbol{r} = &\: \boldsymbol{r}_{\mathrm{e}}-\boldsymbol{r}_{\mathrm{h}},\\
\displaybreak[1]
\nonumber \\
\boldsymbol{R} = &\: \left(m_{\mathrm{e}}\boldsymbol{r}_{\mathrm{e}}+m_{\mathrm{h}}\boldsymbol{r}_{\mathrm{h}}\right)/\left(m_{\mathrm{e}}+m_{\mathrm{h}}\right),\\
\displaybreak[1]
\nonumber \\
\boldsymbol{p} = &\: \left(m_{\mathrm{h}}\boldsymbol{p}_{\mathrm{e}}-m_{\mathrm{e}}\boldsymbol{p}_{\mathrm{h}}\right)/\left(m_{\mathrm{e}}+m_{\mathrm{h}}\right),\\
\displaybreak[1]
\nonumber \\
\boldsymbol{P} = &\: \boldsymbol{p}_{\mathrm{e}}+\boldsymbol{p}_{\mathrm{h}}=\hbar\boldsymbol{K},
\end{align}
\end{subequations}
in Eq.~(\ref{eq:Hpeph}) leads to a coupling term between these motions in
the kinetic energy~\cite{17,24}:
\begin{equation}
H=T_{\mathrm{r}}\left(\boldsymbol{p}\right)+T_{\mathrm{c}}\left(\boldsymbol{p},\,\boldsymbol{K}\right)+T_{\mathrm{t}}\left(\boldsymbol{K}\right)+V\left(\boldsymbol{r}\right).
\end{equation}
In Ref.~\cite{24} a different transformation of coordinates was proposed,
by which the coupling term $T_{\mathrm{c}}$ vanishes. This transformation
reads
\begin{subequations}
\begin{eqnarray}
\left(\boldsymbol{p}_{\mathrm{h}}\right)_{i} & = & \sum_{j}\boldsymbol{A}_{ij}K_{j}-p_{i},\label{eq:phi}\\
\nonumber \\
\left(\boldsymbol{p}_{\mathrm{e}}\right)_{i} & = & \sum_{j}\left(\hbar\delta_{ij}\boldsymbol{1}-\boldsymbol{A}_{ij}\right)K_{j}+p_{i},\label{eq:pei}
\end{eqnarray}
\end{subequations}
where the terms $\boldsymbol{A}_{ij}$ are assumed to be spin matrices.
However, in the calculations of Ref.~\cite{24} the spin-orbit coupling
was assumed to be infinitely large so that only states with $J=3/2$
were considered. We will now calculate the appropriate matrices $\boldsymbol{A}_{ij}$
for the Hamiltonian~(\ref{eq:Hpeph}) and compare the resulting
expression for $T_{\mathrm{t}}\left(\boldsymbol{K}\right)$ with the
ansatz for the exchange interaction in Refs.~\cite{9_1,8,9}.

We define the matrices
\begin{equation}
\boldsymbol{I}_{ij}=3\left\{ \boldsymbol{I}_{i},\,\boldsymbol{I}_{j}\right\} -2\hbar^{2}\delta_{ij}\boldsymbol{1}\label{eq:imatrix}
\end{equation}
according to~\cite{7_11} and note that these operators 
form a closed subset with respect to
the symmetric product $\left\{ a,b\right\} =\frac{1}{2}\left(ab+ba\right)$
(see Appendix~\ref{sub:The-matrices-}). Therefore, we make the
ansatz 
\begin{subequations}
\begin{eqnarray}
\boldsymbol{A}_{jj} & = & \hbar^{2}C_{1}\boldsymbol{1}+\frac{1}{3}C_{2}\boldsymbol{I}_{jj}+\frac{1}{3}C_{3}\boldsymbol{I}_{kl},\\
\nonumber \\
\boldsymbol{A}_{jk} & = & \hbar^{2}C_{4}\boldsymbol{1}+\frac{1}{3}C_{5}\boldsymbol{I}_{jk}+\frac{1}{3}C_{6}\boldsymbol{I}_{ll},
\end{eqnarray}
\end{subequations}
with $j\neq l\neq k\neq j$, which respects the cubic symmetry of
the solid. Inserting Eqs.~(\ref{eq:phi}) and (\ref{eq:pei}) into
the kinetic part of Eq.~(\ref{eq:Hpeph}) and setting the coupling
term $T_{\mathrm{c}}=0$, we can determine the coefficients $C_{i}$.
The $K$-dependent part of the kinetic energy is then exactly of the
same form as the exchange interaction terms in Refs.~\cite{9_1,8,9}:
\begin{eqnarray}
T_{\mathrm{t}}\left(\boldsymbol{K}\right) & = & \Omega_{1}K^{2}\boldsymbol{1}-\frac{\Omega_{3}}{\hbar^2}\left(K_{1}^{2}\boldsymbol{I}_{11}+\mathrm{c.p.}\right)\nonumber \\
 &  & -\frac{2\Omega_{5}}{3\hbar^2}\left(K_{1}K_{2}\boldsymbol{I}_{12}+\mathrm{c.p.}\right)
\end{eqnarray}
or
\begin{widetext}
\begin{equation}
T_{\mathrm{t}}\left(\boldsymbol{K}\right)=\Omega_{1}\left(\begin{array}{ccc}
K^{2} & 0 & 0\\
0 & K^{2} & 0\\
0 & 0 & K^{2}
\end{array}\right)+\Omega_{3}\left(\begin{array}{ccc}
3K_{1}^{2}-K^{2} & 0 & 0\\
0 & 3K_{2}^{2}-K^{2} & 0\\
0 & 0 & 3K_{3}^{2}-K^{2}
\end{array}\right)+\Omega_{5}\left(\begin{array}{ccc}
0 & K_{1}K_{2} & K_{1}K_{3}\\
K_{1}K_{2} & 0 & K_{2}K_{3}\\
K_{1}K_{3} & K_{2}K_{3} & 0
\end{array}\right),
\end{equation}
\label{eq:TtK}
\end{widetext}
where $K$ is now given in units of $k_{0}\approx 2.62\,\mathrm{m}^{-1}$, i.e., the value at the
exciton-photon resonance~\cite{9}. The dependency of the coefficients $C_{i}$
and $\Omega_{i}$ on the Luttinger parameters is given in Appendix~\ref{sub:The-matrices-}.

Since our coefficients $\Omega_i$ cannot be directly compared with the 
according coefficients $\Delta_{i}$ in Refs.~\cite{9_1,8,9}, we use a different symbol
to illustrate this fact.
The impossibility of a direct comparison arises due to three important facts:

First, the operator $T_{\mathrm{t}}\left(\boldsymbol{K}\right)$ describes the kinetic energy 
related to the motion of the center of mass, 
whereas Eq.~(\ref{eq:excans}) only describes the ``exchange interaction,'' i.e.,
the interaction without the spherically symmetric part of the kinetic energy. 
Therefore, it is $\Omega_1\hat{=}\hbar^{2}K^{2}/(2M)+\Delta_1$, $\Omega_3\hat{=}\Delta_3$, and $\Omega_5\hat{=}\Delta_5$
with the exciton mass of the simple-band model $M=m_{\mathrm{e}}+m_{\mathrm{h}}\approx 1.64\,m_0$.

Furthermore, the central-cell
corrections apply for the $1S$ exciton. 
However, the calculations on these corrections have been done within the simple band model 
and the coefficients $c_{\mathrm{e}}=1.35$, $c_{\mathrm{h}}=1.35$ and $d=0.18$, which have been 
introduced in Sec.~\ref{sub:Hamiltonian}, were obtained by comparing results to experimental data.
Effects due to the complex valence band structure may therefore be already included 
in these central-cell corrections so that we cannot separate the true central-cell effects from the effects 
due to $T_{\mathrm{t}}\left(\boldsymbol{K}\right)$.

Finally, since the coefficients $\Delta_i$ were obtained from experimental data,
we have to consider that the $K$-dependent splitting is 
also influenced by the binding energy of the exciton. Therefore,
$T_{\mathrm{t}}\left(\boldsymbol{K}\right)$ 
has to either be included in the matrix diagonalization or at least 
be treated within perturbation theory as was done in Ref.~\cite{24}.

\begin{figure}
\begin{centering}
\includegraphics[width=1.0\columnwidth]{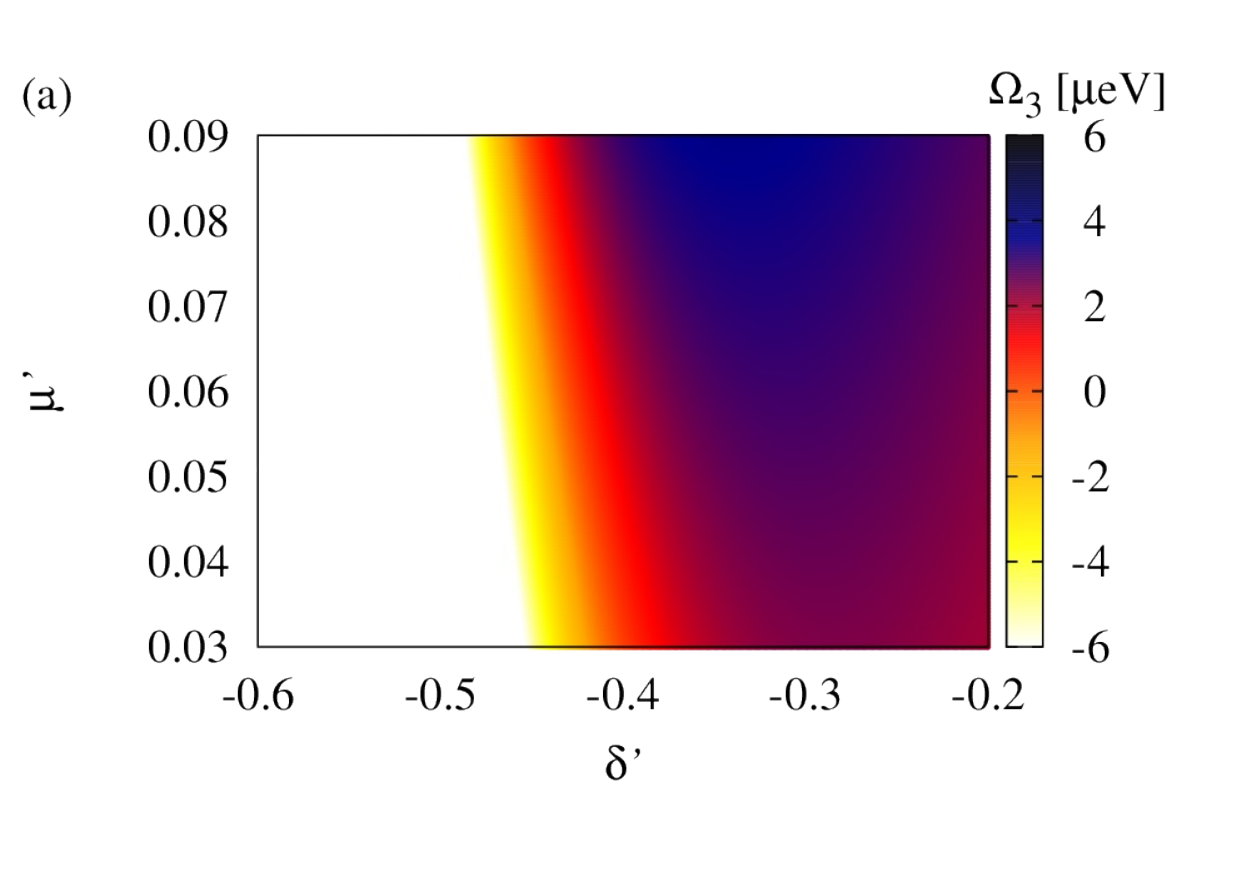}
\includegraphics[width=1.0\columnwidth]{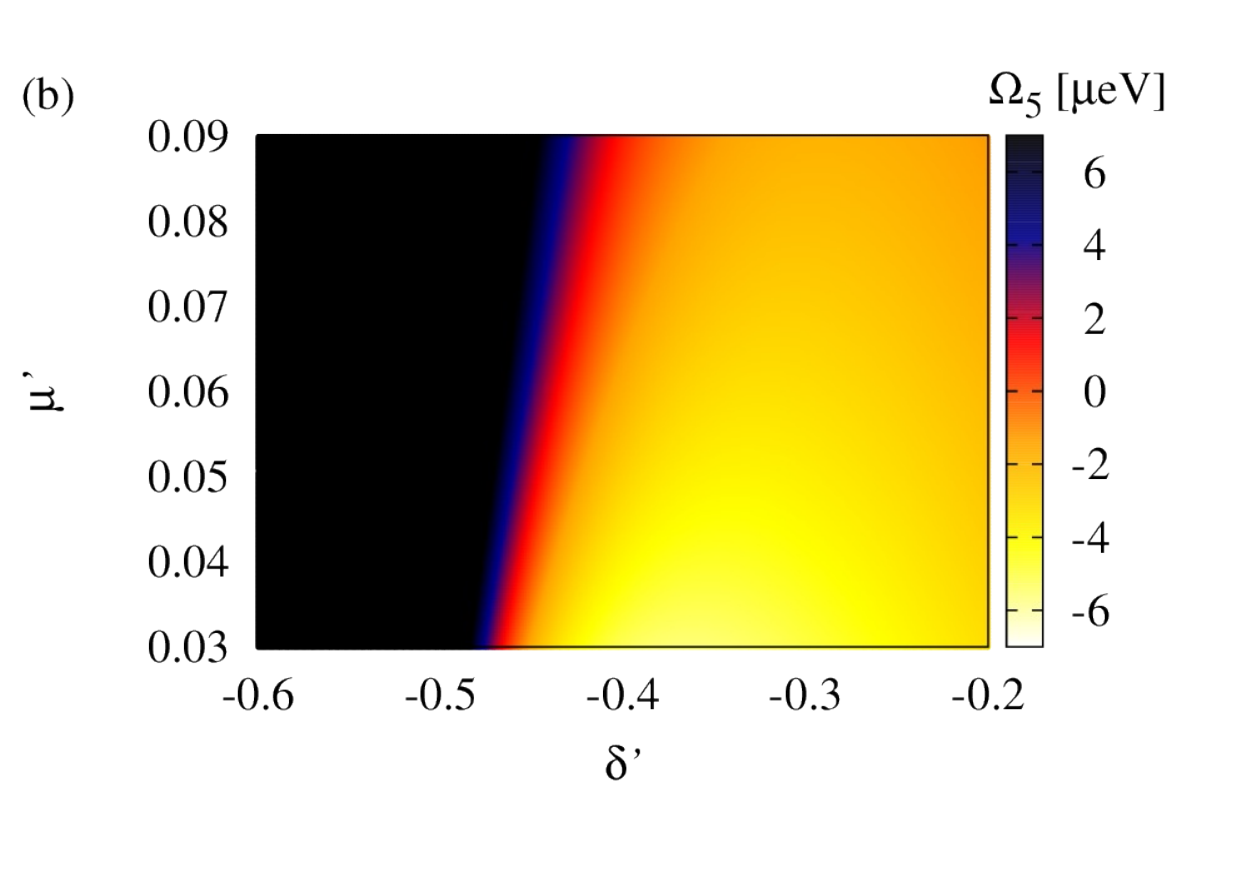}
\par\end{centering}

\protect\caption{(Color online) The values of the coefficients (a) 
$\Omega_3$ and (b) $\Omega_5$ as functions of $\mu'$ and $\delta'$.
\label{fig:8}}
\end{figure}

However, Fig.~\ref{fig:8} shows that there are significant changes in the 
values of the coefficients $\Omega_3$ and $\Omega_5$ if
the parameters $\mu'$ and $\delta'$ are only slightly varied from $\mu'=0.0586$ and $\delta'=-0.404$.
Consequently, a more comprehensive analysis of 
$T_{\mathrm{t}}\left(\boldsymbol{K}\right)$ would not give more reliable results due to
the small uncertainties in $\mu'$ and $\delta'$, which have been 
discussed at the end of Sec.~\ref{sub:Determiantion-of}.

Therefore, we present only a very simple analysis of $T_{\mathrm{t}}\left(\boldsymbol{K}\right)$:
Using $\gamma_{1}'=2.77$ as a given value, we solve the equations 
\begin{subequations}
\begin{eqnarray}
-1.3~\mathrm{\mu eV} & = & \Omega_3(\mu',\delta'),\\
2.0~\mathrm{\mu eV} & = & \Omega_5(\mu',\delta')
\end{eqnarray}
\end{subequations}
for $\mu'$ and $\delta'$ and obtain $\mu'\approx 0.0583$ and $\delta' \approx -0.442$.
Inserting these results in $\Omega_1(\mu',\delta')$ yields
$\Omega_1\approx 2.15~\mathrm{\mu eV}$ or 
$\Omega_1-\hbar^{2}K^{2}/(2M)\approx -13.73~\mathrm{\mu eV}$.
Of course, the deviation from the experimental value $\Delta_1=-8.6~\mathrm{\mu eV}$
could now be explained in terms of the simplicity of this analysis.

However, we have proved that there is clear evidence that
the valence band structure or the term 
$T_{\mathrm{t}}\left(\boldsymbol{K}\right)$ describing the kinetic energy 
of the motion of the center of mass of the exciton
is the cause for the $K$-dependent behavior of 
the $1S$ exciton observed in Refs.~\cite{9_1,8,9}, whereas a $K$-dependent short-range interaction
can most likely be excluded~\cite{1}.

\section{Summary and outlook\label{sec:Summary-and-outlook}}

Using the method of solving the Schr\"odinger equation in a complete
basis, we determined the eigenstates of the cubic Hamiltonian of excitons
accounting for the complex band structure. An evaluation of the calculated
line spacings and oscillator strengths proved that there are several combinations of 
the parameters $\mu'$ and $\delta'$ by which the excitonic spectra can be described. 
However, we assume that the values resulting from band structure calculations are the correct ones.
Using the values $\gamma'_{1}=2.77$, $\mu'=0.0586$, and $\delta'=-0.404$, 
the Luttinger parameters of $\mathrm{Cu_{2}O}$ are
\begin{equation}
\gamma_{1}=1.76,\quad\gamma_{2}\approx0.75,\quad\gamma_{3}\approx-0.37.\label{eq:gammai}
\end{equation}
Furthermore,
we separated the relative motion and the motion of the center of mass
using an appropriate coordinate transformation. The final result allows
us to explain the $K$-dependent line splitting of the $1S$ exciton
state in terms of the complex kinetic energy of the motion of the center of mass. 
As a next step,
we plan to extend our method to calculate excitonic spectra of $\mathrm{Cu_{2}O}$ in the
presence of external magnetic and electric fields.

\acknowledgments
We thank M.~M. Glazov, M.~A. Semina, D. Fr\"ohlich, M. Bayer and H. Cartarius for helpful discussions.

\appendix

\section{Oscillator strengths \label{sub:Oscillator-strengths}}

We now give the formula for the relative oscillator strength (see Sec.~\ref{sub:Oscillator-strengths-1})
\begin{equation}
f_{\mathrm{rel}}\sim\left|\lim_{r\rightarrow0}\frac{\partial}{\partial r}\left\langle D\middle|\Psi\left(\boldsymbol{r}\right)\right\rangle\right|^2.
\end{equation}
Using the wave function of Eq.~(\ref{eq:ansatz}), we find
\begin{eqnarray}
f_{\mathrm{rel}} & \sim & \left|\sum_{NJFF_{t}}\sum_{M_{S_{\mathrm{e}}}M_{I}}\sum_{k=\pm2}c_{N1JFF_{t}k}\right.\nonumber \\
\nonumber \\
 &  & \times k\,\left(-1\right)^{F-J-3M_{S_{\mathrm{e}}}-M_{I}+\frac{3}{2}}\left[(N+1)(N+3)\right]^{\frac{1}{2}}\nonumber \\
\nonumber \\
 &  & \times\left[(2J+1)(2F+1)(2F_{t}+1)\right]^{\frac{1}{2}}\nonumber \\
\nonumber \\
 &  & \times\left(\begin{array}{ccc}
1 & \frac{1}{2} & J\\
M_{I} & -M_{S_{\mathrm{e}}} & M_{S_{\mathrm{e}}}-M_{I}
\end{array}\right)\nonumber \\
\nonumber \\
 &  & \times\left(\begin{array}{ccc}
1 & 1 & 2\\
M_{I} & k-M_{I} & -k
\end{array}\right)\nonumber \\
\nonumber \\
 &  & \times\left(\begin{array}{ccc}
F & \frac{1}{2} & F_{t}\\
k-M_{S_{\mathrm{e}}} & M_{S_{\mathrm{e}}} & -k
\end{array}\right)\nonumber \\
\nonumber \\
 &  & \left.\times\left(\begin{array}{ccc}
1 & J & F\\
k-M_{I} & M_{I}-M_{S_{\mathrm{e}}} & M_{S_{\mathrm{e}}}-k
\end{array}\right)\right|^{2}.
\end{eqnarray}

\section{Recursion relations of the Coulomb-Sturmian functions \label{sub:Recursion-relations-of}}

In this section we give all important recursion relations of the Coulomb-Sturmian
 functions based on the calculations in Ref.~\cite{S2}.
In this regard, we also give the recursion relations needed if external
magnetic or electric fields are present. 
The Coulomb-Sturmian functions read
\begin{eqnarray}
\phi_{N,L,M}\left(\boldsymbol{r}\right) & = & U_{NL}\left(\rho\right)Y_{LM}\left(\Omega\right)\nonumber \\
\nonumber \\
 & = & N_{NL}\left(2\rho\right)^{L}e^{-\rho}L_{N}^{2L+1}\left(2\rho\right)Y_{LM}\left(\Omega\right)\quad
\end{eqnarray}
with $\rho=r/\alpha$, a normalization factor
\begin{equation}
N_{NL}=\frac{2}{\sqrt{\alpha^{3}}}\left[\frac{N!}{\left(N+2L+1\right)!\left(N+L+1\right)}\right]^{\frac{1}{2}},
\end{equation}
the associated Laguerre polynomials $L_{n}^{m}\left(x\right)$, and
an arbitrary scaling parameter $\alpha$.
The radial functions 
obey the orthogonality relation
\begin{equation}
\int_{0}^{\infty}\,\mathrm{d}r\; rU_{N'L}\left(r\right)U_{NL}\left(r\right)=\frac{1}{\alpha\left(N+L+1\right)}\delta_{NN'}.
\end{equation}

For the recursion relations we set $\alpha=1$
and omit the dependency of the functions $\phi_{N,L,M}$
on $\boldsymbol{r}$. Coefficients in these relations will be given,
e.g., in the form $\left(R_{1}\right)_{NL}^{j}$, which means that
they are functions of $j$, $n$, and $L$, etc. The vectors $\hat{n}$
and $\nabla^{\hat{n}}$ are defined as in Ref.~\cite{S2}.


\begin{align}
\hat{n}_{3}\phi_{N,L,M}= &\: \sum_{j=\pm1}\left(N_{1}\right)_{LM}^{j}\phi_{N,L+j,M}
\\
\displaybreak[1]
\nonumber \\
\left(N_{1}\right)_{LM}^{j}= &\: \delta_{1j}\left\{ \left[\frac{\left(L+M+1\right)\left(L-M+1\right)}{\left(2L+1\right)\left(2L+3\right)}\right]^{\frac{1}{2}}\right\}
\nonumber \\
\nonumber \\
+ &\: \delta_{-1j}\left\{ \left[\frac{\left(L+M\right)\left(L-M\right)}{\left(2L+1\right)\left(2L-1\right)}\right]^{\frac{1}{2}}\right\} 
\displaybreak[1]
\\
\nonumber \\
\nabla_{3}^{\hat{n}}\phi_{N,L,M}= &\: \sum_{j=\pm1}\left(D_{1}\right)_{LM}^{j}\phi_{N,L+j,M}
\\
\displaybreak[1]
\nonumber \\
\left(D_{1}\right)_{LM}^{j}= &\: \delta_{1j}\left\{ -L\left(N_1\right)_{LM}^{1}\right\}\nonumber \\
\nonumber \\
+ &\: \delta_{-1j}\left\{ \left(L+1\right)\left(N_1\right)_{LM}^{-1}\right\} 
\\
\displaybreak[1]
\nonumber \\
r\phi_{N,L,M}= &\: \sum_{j=-1}^{1}\left(R_{1}\right)_{NL}^{j}\phi_{N+j,L,M}
\\
\displaybreak[1]
\nonumber \\
\left(R_{1}\right)_{NL}^{j}= &\: \delta_{1j}\left\{ -\frac{1}{2}\left[\left(N+1\right)\left(N+L+2\right)\right]^{\frac{1}{2}}\right.\nonumber \\
 &\: \qquad\qquad\left.\times\left[\frac{\left(N+2L+2\right)}{\left(N+L+1\right)}\right]^{\frac{1}{2}}\right\} \nonumber \\
\nonumber \\
+ &\: \delta_{0j}\left\{ N+L+1\right\} \nonumber \\
\nonumber \\
+ &\: \delta_{-1j}\left\{ -\frac{1}{2}\left[\left(N\right)\left(N+L\right)\right]^{\frac{1}{2}}\right.\nonumber \\
 &\: \qquad\qquad\left.\times\left[\frac{\left(N+2L+1\right)}{\left(N+L+1\right)}\right]^{\frac{1}{2}}\right\} 
\\
\displaybreak[1]
\nonumber \\
r\frac{\partial}{\partial r}\phi_{N,L,M}= &\: \sum_{j=-1}^{-1}\left(RP_{1}\right)_{NL}^{j}\phi_{N+j,L,M}
\\
\displaybreak[1]
\nonumber \\
\left(RP_{1}\right)_{NL}^{j}= &\: \delta_{-1j}\left\{ \left(R_{1}\right)_{NL}^{-1}\right\}\nonumber \\
\nonumber \\
+ &\: \delta_{0j}\left\{ -1\right\}+\delta_{1j}\left\{ -\left(R_{1}\right)_{NL}^{1}\right\} 
\\
\displaybreak[1]
\nonumber \\
r\phi_{N,L,M}= &\: \sum_{j=-2}^{0}\left(L_{1}\right)_{NL}^{j\,1}\phi_{N+j,L+1,M}
\nonumber \\
= &\: \sum_{j=0}^{2}\left(L_{1}\right)_{NL}^{j\,-1}\phi_{N+j,L-1,M}
\\
\displaybreak[1]
\nonumber \\
\left(L_{1}\right)_{NL}^{j\, k}= &\: \delta_{2j}\delta_{-1k}\left\{ \frac{1}{2}\left[\left(N+2\right)\left(N+1\right)\right]^{\frac{1}{2}}\right.\nonumber \\
 &\: \qquad\qquad\times\left.\left[\frac{\left(N+L+2\right)}{\left(N+L+1\right)}\right]^{\frac{1}{2}}\right\} \nonumber \\
\nonumber \\
+ &\: \delta_{1j}\delta_{-1k}\left\{ -\left[\left(N+1\right)\left(N+2L+1\right)\right]^{\frac{1}{2}}\right\} \nonumber \\
\nonumber \\
+ &\: \delta_{0j}\delta_{-1k}\left\{ \frac{1}{2}\left[\left(N+2L\right)\left(N+2L+1\right)\right]^{\frac{1}{2}}\right. \nonumber \\
 &\: \qquad\qquad\times\left.\left[\frac{\left(N+L\right)}{\left(N+L+1\right)}\right]^{\frac{1}{2}}\right\} \nonumber \\
\nonumber \\
+ &\: \delta_{0j}\delta_{1k}\left\{ \frac{1}{2}\left[\left(N+2L+2\right)\left(N+2L+3\right)\right]^{\frac{1}{2}}\right. \nonumber \\
 &\: \qquad\qquad\times\left.\left[\frac{\left(N+L+2\right)}{\left(N+L+1\right)}\right]^{\frac{1}{2}}\right\} \nonumber \\
\nonumber \\
+ &\: \delta_{-1j}\delta_{1k}\left\{ -\left[\left(N\right)\left(N+2L+2\right)\right]^{\frac{1}{2}}\right\} \nonumber \\
\nonumber \\
+ &\: \delta_{-2j}\delta_{1k}\left\{ \frac{1}{2}\left[\left(N\right)\left(N-1\right)\right]^{\frac{1}{2}}\right. \nonumber \\
 &\: \qquad\qquad\times\left.\left[\frac{\left(N+L\right)}{\left(N+L+1\right)}\right]^{\frac{1}{2}}\right\} 
\end{align}



With these relations we calculate combined formulas:
\begin{align}
r^{2}\phi_{N,L,M}= &\: \sum_{j=-2}^{2}\left(R_{2}\right)_{NL}^{j}\phi_{N+j,L,M}
\\
\displaybreak[1]
\nonumber \\
\left(R_{2}\right)_{NL}^{j}= &\: \sum_{w=j-1}^{j+1}\left(R_{1}\right)_{NL}^{w}\left(R_{1}\right)_{N+w\, L}^{j-w}
\\
\displaybreak[1]
\nonumber \\
r^{3}\phi_{N,L,M}= &\: \sum_{j=-3}^{3}\left(R_{3}\right)_{NL}^{j}\phi_{N+j,L,M}
\\
\displaybreak[1]
\nonumber \\
\left(R_{3}\right)_{NL}^{j}= &\: \sum_{w=j-1}^{j+1}\left(R_{2}\right)_{NL}^{w}\left(R_{1}\right)_{N+w\, L}^{j-w}
\\
\displaybreak[1]
\nonumber \\
r\hat{n}_{3}\phi_{N,L,M}= &\: \sum_{k=\pm1}\,\sum_{j=-1-k}^{1-k}\left(LN_{1}\right)_{NLM}^{j\, k}
\nonumber \\
\times &\: \phi_{N+j,L+k,M}
\\
\displaybreak[1]
\nonumber \\
\left(LN_{1}\right)_{NLM}^{j\, k}= &\: \left(L_{1}\right)_{NL}^{j\, k}\left(N_{1}\right)_{LM}^{k}
\\
\displaybreak[1]
\nonumber \\
r^{2}\hat{n}_{3}\phi_{N,L,M}= &\: \sum_{k=\pm1}\,\sum_{j=-2-k}^{2-k}\left(RLN_{1}\right)_{NLM}^{j\, k}\nonumber \\
\nonumber \\
\times &\: \phi_{N+j,L+k,M}
\\
\displaybreak[1]
\nonumber \\
\left(RLN_{1}\right)_{NLM}^{j\, k}= &\: \sum_{w=j-1}^{j+1}\left(LN_{1}\right)_{NLM}^{w\, k}\nonumber \\
\nonumber \\
\times &\: \left(R_{1}\right)_{N+w\, L+k}^{j-w}
\\
\displaybreak[1]
\nonumber \\
r^{2}\hat{n}_{3}^{2}\phi_{N,L,M}= &\: \sum_{k=0,\pm2}\,\sum_{j=-2-k}^{2-k}\left(LN_{2}\right)_{NLM}^{j\, k}\nonumber \\
\nonumber \\
\times &\: \phi_{N+j,L+k,M}
\\
\displaybreak[1]
\nonumber \\
\left(LN_{2}\right)_{NLM}^{j\, k}= &\: \sum_{v=\pm1}\,\sum_{w=j-1-v}^{j+1-v}\left(LN_{1}\right)_{NLM}^{w\, k+v}\nonumber \\
\nonumber \\
\times &\: \left(LN_{1}\right)_{N+w\, L+k+v\, M}^{j-w\,-v}
\\
\displaybreak[1]
\nonumber \\
r^{3}\hat{n}_{3}^{2}\phi_{N,L,M}= &\: \sum_{k=0,\pm2}\,\sum_{j=-3-k}^{3-k}\left(R_{1}LN_{2}\right)_{NLM}^{j\, k}\nonumber \\
\nonumber \\
\times &\: \phi_{N+j,L+k,M}
\\
\displaybreak[1]
\nonumber \\
\left(R_{1}LN_{2}\right)_{NLM}^{j\, k}= &\: \sum_{w=j-1}^{j+1}\left(LN_{2}\right)_{NLM}^{w\, k}\nonumber \\
\nonumber \\
\times &\: \left(R_{1}\right)_{N+w\, L+k}^{j-w}
\\
\displaybreak[1]
\nonumber \\
r^{2}\frac{\partial}{\partial r}\phi_{N,L,M}= &\: \sum_{j=-2}^{2}\left(R_{2}P_{1}\right)_{NL}^{j}\phi_{N+j,L,M}
\\
\displaybreak[1]
\nonumber \\
\left(R_{2}P_{1}\right)_{NL}^{j}= &\: \sum_{w=j-1}^{j+1}\left(RP_{1}\right)_{NL}^{w}\left(R_{1}\right)_{N+w\, L}^{j-w}
\\
\displaybreak[1]
\nonumber \\
r^{3}\frac{\partial}{\partial r}\phi_{N,L,M}= &\: \sum_{j=-3}^{3}\left(R_{3}P_{1}\right)_{NL}^{j}\phi_{N+j,L,M}
\\
\displaybreak[1]
\nonumber \\
\left(R_{3}P_{1}\right)_{NL}^{j}= &\: \sum_{w=j-1}^{j+1}\left(R_{2}P_{1}\right)_{NL}^{w}\left(R_{1}\right)_{N+w\, L}^{j-w}
\end{align}

\section{Matrix elements \label{sub:Matrix-elements}}

In this section we give all matrix elements of the terms of the Hamiltonian
$H$ [Eq.~\ref{eq:H0}] in the basis of Eq.~(\ref{eq:basis})
in Hartree units using the formalism of irreducible tensors~\cite{ED}.
We use the abbreviation
\begin{equation}
\tilde{\delta}_{\Pi\Pi'}=\delta_{LL'}\delta_{JJ'}\delta_{FF'}\delta_{F_{t}F'_{t}}\delta_{M_{F_{t}}M'_{F_{t}}}
\end{equation}
in the following. The functions of the form $\left(R_{1}\right)_{NL}^{j}$
are taken from the recursion relations of the Coulomb-Sturmian functions
in Appendix~\ref{sub:Recursion-relations-of}. The value of the integral
$I_{N'\, L';N\, L}$ is given in Appendix~\ref{sub:Reduced-matrix-elements}.

\begin{widetext}

\begin{align}
\left\langle \Pi'\left|r^{-1}\right|\Pi\right\rangle = &\: \tilde{\delta}_{\Pi\Pi'}\delta_{NN'}\left[N+L+1\right]^{-1}
\\
\displaybreak[1]
\nonumber \\
\left\langle \Pi'\left|1\right|\Pi\right\rangle = &\: \tilde{\delta}_{\Pi\Pi'}\sum_{j=-1}^{1}\left(R_{1}\right)_{NL}^{j}\left[N+L+j+1\right]^{-1}\delta_{N',N+j}
\\
\displaybreak[1]
\nonumber \\
\left\langle \Pi'\left|p^{2}\right|\Pi\right\rangle = &\: 2\tilde{\delta}_{\Pi\Pi'}\delta_{NN'}-\tilde{\delta}_{\Pi\Pi'}\sum_{j=-1}^{1}\left(R_{1}\right)_{NL}^{j}\left[N+L+j+1\right]^{-1}\delta_{N',N+j}
\\
\displaybreak[1]
\nonumber \\
\left\langle \Pi'\left|1+I^{(1)}\cdot S_{\mathrm{h}}^{(1)}\right|\Pi\right\rangle = &\: \tilde{\delta}_{\Pi\Pi'}\left(\frac{1}{2}J\left(J+1\right)-\frac{3}{8}\right)\sum_{j=-1}^{1}\left(R_{1}\right)_{NL}^{j}\left[N+L+j+1\right]^{-1}\delta_{N',N+j}
\\
\displaybreak[1]
\nonumber \\
\left\langle \Pi'\left|P^{(2)}\cdot I^{(2)}\right|\Pi\right\rangle = &\: \sqrt{5}\,\left\langle \Pi'\left|\left[P^{(2)}\times I^{(2)}\right]_{0}^{(0)}\right|\Pi\right\rangle \label{eq:PI}
\\
\displaybreak[1]
\nonumber \\
\left\langle \Pi'\left|\left[P^{(2)}\times I^{(2)}\right]_{q}^{(K)}\right|\Pi\right\rangle = &\: 3\sqrt{5}\left(-1\right)^{F'_{t}+F_{t}-M'_{F_{t}}+F'+J+K}\left\langle N'\, L'\left\Vert P^{(2)}\right\Vert N\, L\right\rangle\nonumber \\
\nonumber \\
\times &\: \left[\left(2F_{t}+1\right)\left(2F'_{t}+1\right)\left(2F+1\right)\left(2F'+1\right)\left(2K+1\right)\left(2J+1\right)\left(2J'+1\right)\right]^{\frac{1}{2}}\nonumber \\
\nonumber \\
\times &\: \left(\begin{array}{ccc}
F'_{t} & K & F_{t}\\
-M'_{F_{t}} & q & M_{F_{t}}
\end{array}\right)\left\{ \begin{array}{ccc}
F' & F'_{t} & \frac{1}{2}\\
F_{t} & F & K
\end{array}\right\}\left\{ \begin{array}{ccc}
1 & J' & \frac{1}{2}\\
J & 1 & 2
\end{array}\right\}\left\{ \begin{array}{ccc}
L' & L & 2\\
J' & J & 2\\
F' & F & K
\end{array}\right\}\label{eq:9j13}
\\
\displaybreak[1]
\nonumber \\
\left\langle \Pi'\left|p^{2}\left(I^{(1)}\cdot S_{\mathrm{h}}^{(1)}\right)\right|\Pi\right\rangle = &\: \frac{1}{2}\left(J\left(J+1\right)-\frac{11}{4}\right)\left\langle \Pi'\left|p^{2}\right|\Pi\right\rangle
\\
\displaybreak[1]
\nonumber \\
\left\langle \Pi'\left|P^{(2)}\cdot\left[I^{(2)}\times S_{\mathrm{h}}^{(1)}\right]^{(2)}\right|\Pi\right\rangle = &\: \sqrt{5}\left\langle \Pi'\left|\left[P^{(2)}\times\left[I^{(2)}\times S_{\mathrm{h}}^{(1)}\right]^{(2)}\right]_{0}^{(0)}\right|\Pi\right\rangle 
\\
\displaybreak[1]
\nonumber \\
\left\langle \Pi'\left|\left[P^{(2)}\times\left[I^{(2)}\times S_{\mathrm{h}}^{(1)}\right]^{(2)}\right]_{q}^{(K)}\right|\Pi\right\rangle = &\:  3\sqrt{5}\left(-1\right)^{F_{t}'+F_{t}-M_{F_{t}}'+F'+K+\frac{1}{2}}\left\langle N\, L'\left\Vert P^{(2)}\right\Vert N\, L\right\rangle \nonumber \\
\nonumber \\
\times &\:[(2F_{t}+1)(2F_{t}'+1)(2F+1)(2F'+1)(2K+1)(2J+1)(2J'+1)]^{\frac{1}{2}}\nonumber \\
\nonumber \\
\times &\:\left(\begin{array}{ccc}
F_{t}' & K & F_{t}\\
-M_{F_{t}}' & q & M_{F_{t}}
\end{array}\right)\left\{ \begin{array}{ccc}
F' & F_{t}' & \frac{1}{2}\\
F_{t} & F & K
\end{array}\right\} \left\{ \begin{array}{ccc}
L' & L & 2\\
J' & J & 2\\
F' & F & K
\end{array}\right\} \left\{ \begin{array}{ccc}
1 & 1 & 1\\
\frac{1}{2} & \frac{1}{2} & 1\\
J' & J & 2
\end{array}\right\} 
\end{align}

\end{widetext}

In Eq.~(\ref{eq:9j13}) the order of the quantum numbers in the rows of the 
$9j$-symbol is given by relations in Refs.~\cite{ED,86}
and differs from Eq.~(14) of the Supplementary Material of Ref.~\cite{28} 
\textcolor{black}{or Eq.~(A2) of Ref.~\cite{17_17_26}.
This odd permutation of rows changes the sign of the sign of the 
$9j$-symbol by~\cite{ED}
\begin{equation}
(-1)^{L'+L+2+J'+J+2+F'+F+4}.
\end{equation}
So there is change in the sign of the $9j$-symbol, e.g., 
for $L=L'=3$, $J=1/2$, $J'=3/2$ and $F=F'=5/2$ (cf. Eq.~(15a) 
of the Supplementary Material of Ref.~\cite{28})}.

\section{Reduced matrix elements \label{sub:Reduced-matrix-elements}}

We now give the value of the reduced matrix element $\left\langle N'\, L'\left\Vert P^{(2)}\right\Vert N\, L\right\rangle $.
We use the abbreviation $I_{N'\, L';N\, L}$ for the integral
\begin{eqnarray}
I_{N'\, L';N\, L} & = & \int_{0}^{\infty}\,\mathrm{d}r\, U_{N'L'}\left(r\right)U_{NL}\left(r\right)\nonumber \\
\nonumber \\
 & = & 2\left[\frac{N'!N!\left(N'+2L'+1\right)!\left(N+2L+1\right)!}{\left(N'+L'+1\right)\left(N+L+1\right)}\right]^{\frac{1}{2}}\nonumber \\
\nonumber \\
 &  & \times\sum_{k=0}^{N'}\sum_{j=0}^{N}\frac{\left(k+j+L+L'\right)!}{k!j!\left(N'-k\right)!\left(N-j\right)!}\nonumber \\
\nonumber \\
 &  & \qquad\quad\times\frac{\left(-1\right)^{k+j}}{\left(k+2L'+1\right)!\left(j+2L+1\right)!}
\end{eqnarray}
with $\alpha=1$. The functions of the form $\left(R_{1}\right)_{NL}^{j}$
are taken from the recursion relations of the Coulomb-Sturmian functions
in Appendix~\ref{sub:Recursion-relations-of}.

\begin{widetext}

\begin{eqnarray}
\left\langle N'\, L'\left\Vert P^{(2)}\right\Vert N\, L\right\rangle  & = & \delta_{L',L+2}\;\frac{3}{2}\left[\frac{\left(2L+4\right)\left(2L+2\right)}{\left(2L+3\right)}\right]^{\frac{1}{2}}\nonumber \\
\nonumber \\
 &  & \quad\times\left[\sum_{j=-2}^{2}\left(-\left(R_{2}\right)_{NL}^{j}-\delta_{j0}L\left(2L+3\right)\right)I_{N'\, L+2;\, N+j\, L}\right.\nonumber \\
\nonumber \\
 &  & \quad\qquad+\left.\sum_{j=-1}^{1}\left(\left(2L+3\right)\left(RP_{1}\right)_{NL}^{j}+2\left(N+L+1\right)\left(R_{1}\right)_{NL}^{j}\right)I_{N'\, L+2;\, N+j\, L}\right]\nonumber \\
\nonumber \\
 & + & \delta_{L',L}\;\left(-\sqrt{3}\right)\left[\frac{L\left(2L+1\right)\left(2L+2\right)}{\left(2L+3\right)\left(2L-1\right)}\right]^{\frac{1}{2}}\nonumber \\
\nonumber \\
 &  & \quad\times\left[2\delta_{NN'}-\sum_{j=-1}^{1}\left(R_{1}\right)_{NL}^{j}\left[N+L+j+1\right]^{-1}\delta_{N',N+j}\right]\nonumber \\
\nonumber \\
 & + & \delta_{L',L-2}\;\frac{3}{2}\left[\frac{\left(2L\right)\left(2L-2\right)}{\left(2L-1\right)}\right]^{\frac{1}{2}}\nonumber \\
\nonumber \\
 &  & \quad\times\left[\sum_{j=-2}^{2}\left(-\left(R_{2}\right)_{NL}^{j}+\delta_{j0}\left(1-L-2L^{2}\right)\right)I_{N'\, L-2;\, N+j\, L}\right.\nonumber \\
\nonumber \\
 &  & \quad\qquad+\left.\sum_{j=-1}^{1}\left(\left(1-2L\right)\left(RP_{1}\right)_{NL}^{j}+2\left(N+L+1\right)\left(R_{1}\right)_{NL}^{j}\right)I_{N'\, L-2;\, N+j\, L}\right]
\end{eqnarray}

\end{widetext}

\section{The matrices $\boldsymbol{I}_{jk}$ \label{sub:The-matrices-}}

In Sec.~\ref{sub:-dependent-line-splitting} we introduced the
matrices
\begin{equation}
\boldsymbol{I}_{ij}=3\left\{ \boldsymbol{I}_{i},\,\boldsymbol{I}_{j}\right\} -2\hbar^{2}\delta_{ij}\boldsymbol{1}.\label{eq:imatrix-1}
\end{equation}
We will shortly list the main properties of these matrices. The second-rank
tensor with the components $\boldsymbol{I}_{ij}$ is symmetric
\begin{equation}
\boldsymbol{I}_{ij}=\boldsymbol{I}_{ji}
\end{equation}
and traceless 
\begin{equation}
\sum_{i=1}^{3}\boldsymbol{I}_{ii}=\boldsymbol{0}.
\end{equation}
As already stated in Sec.~\ref{sub:-dependent-line-splitting}
the operators $\boldsymbol{I}_{ij}$ form a closed subset with respect
to the symmetric product $\left\{ a,b\right\} =\frac{1}{2}\left(ab+ba\right)$:

\begin{subequations}
\begin{align}
\left\{ \boldsymbol{I}_{jj},\,\boldsymbol{I}_{jj}\right\} = &\: \hbar^{2}\left(2\hbar^{2}\boldsymbol{1}-\boldsymbol{I}_{jj}\right),\\
\displaybreak[1]
\nonumber \\
\left\{ \boldsymbol{I}_{jk},\,\boldsymbol{I}_{jk}\right\} = &\: -\frac{3\hbar^{2}}{4}\left(-2\hbar^{2}\boldsymbol{1}+\boldsymbol{I}_{jj}+\boldsymbol{I}_{kk}\right),\\
\displaybreak[1]
\nonumber \\
\left\{ \boldsymbol{I}_{jj},\,\boldsymbol{I}_{kk}\right\} = &\: \hbar^{2}\left(-\hbar^{2}\boldsymbol{1}+\boldsymbol{I}_{jj}+\boldsymbol{I}_{kk}\right),\\
\displaybreak[1]
\nonumber \\
\left\{ \boldsymbol{I}_{jj},\,\boldsymbol{I}_{jk}\right\} = &\: -\frac{\hbar^{2}}{2}\boldsymbol{I}_{jk},\\
\displaybreak[1]
\nonumber \\
\left\{ \boldsymbol{I}_{jk},\,\boldsymbol{I}_{kl}\right\} = &\: -\frac{3\hbar^{2}}{4}\boldsymbol{I}_{jl},\\
\displaybreak[1]
\nonumber \\
\left\{ \boldsymbol{I}_{jj},\,\boldsymbol{I}_{kl}\right\} = &\: \hbar^{2}\boldsymbol{I}_{kl},
\end{align}
\end{subequations}
where $j\neq l\neq k\neq j$.

The matrices can also be expressed in terms of irreducible tensors~\cite{7_11}:
\begin{subequations}
\begin{align}
\boldsymbol{I}_{11}  = &\: \frac{1}{2}\left[I_{2}^{(2)}+I_{-2}^{(2)}-\sqrt{\frac{2}{3}}I_{0}^{(2)}\right]\\
\displaybreak[1]
\nonumber \\
\boldsymbol{I}_{22}  = &\: -\frac{1}{2}\left[I_{2}^{(2)}+I_{-2}^{(2)}+\sqrt{\frac{2}{3}}I_{0}^{(2)}\right]\\
\displaybreak[1]
\nonumber \\
\boldsymbol{I}_{33}  = &\: \sqrt{\frac{2}{3}}I_{0}^{(2)}\\
\displaybreak[1]
\nonumber \\
\boldsymbol{I}_{12}  = &\: -\frac{i}{2}\left[I_{2}^{(2)}-I_{-2}^{(2)}\right]\\
\displaybreak[1]
\nonumber \\
\boldsymbol{I}_{23}  = &\: \frac{i}{2}\left[I_{1}^{(2)}+I_{-1}^{(2)}\right]\\
\displaybreak[1]
\nonumber \\
\boldsymbol{I}_{31}  = &\: -\frac{1}{2}\left[I_{1}^{(2)}-I_{-1}^{(2)}\right]
\end{align}
\end{subequations}

Furthermore, we list the results for the coefficients $C_{i}$ and
the dependency of the parameters $\Omega_{i}$ of Sec.~\ref{sub:-dependent-line-splitting}
on the Luttinger parameters. The coefficients read 
\begin{subequations}
\begin{align}
C_{1} = &\: \Xi\cdot\left(2\gamma_{1}^{'2}+4\gamma_{2}^{2}-3\gamma_{3}^{2}+6\gamma'_{1}\gamma_{2}\right.\nonumber \\
 &\: \quad\quad\left.+6\gamma_{2}\gamma_{3}+3\gamma'_{1}\gamma_{3}\right)\\
\displaybreak[1]
\nonumber \\
C_{2} = &\: 6\Xi\cdot\left(2\gamma_{2}^{2}-3\gamma_{3}^{2}+2\gamma'_{1}\gamma_{2}+3\gamma_{2}\gamma_{3}\right)\\
\displaybreak[1]
\nonumber \\
C_{5} = &\: 12\Xi\cdot\left(\gamma_{2}\gamma_{3}+\gamma'_{1}\gamma_{3}\right)\\
\displaybreak[1]
\nonumber \\
C_{3} = &\: C_{4}=C_{6}=0
\end{align}
\end{subequations}
with
\begin{eqnarray}
\Xi & = & \frac{m_{0}}{\hbar m_{\mathrm{e}}}\left[\left(\gamma'_{1}-2\gamma_{2}\right)\left(\gamma'_{1}+4\gamma_{2}\right)\right.\nonumber \\
 &  & \qquad\quad\left.\times\left(2\gamma'_{1}+2\gamma_{2}+3\gamma_{3}\right)-27\gamma'_{1}\gamma_{3}^{2}\right]^{-1}
\end{eqnarray}
The parameters $\Omega_{i}$ are given by
\begin{subequations}
\begin{eqnarray}
\frac{\Omega_{1}}{k_{0}^{2}} & = & \frac{\hbar^{2}}{2m_{\mathrm{e}}}-\frac{\hbar^{3}}{m_{\mathrm{e}}}C_{1}\nonumber \\
\nonumber \\
 &  & +\frac{\hbar^{4}\gamma'_{1}}{18m_{0}}\left(9C_{1}^{2}+2C_{2}^{2}+3C_{5}^{2}\right)\nonumber \\
\nonumber \\
 &  & -\frac{\hbar^{4}\gamma_{2}}{18m_{0}}\left(24C_{1}C_{2}-4C_{2}^{2}-3C_{5}^{2}\right)\nonumber \\
\nonumber \\
 &  & -\frac{\hbar^{4}\gamma_{3}}{12m_{0}}C_{5}\left(24C_{1}-4C_{2}-3C_{5}\right)
\end{eqnarray}

\begin{eqnarray}
\frac{\Omega_{3}}{k_{0}^{2}} & = & \frac{\hbar^{3}}{3m_{\mathrm{e}}}C_{2}\nonumber \\
\nonumber \\
 &  & -\frac{\hbar^{4}\gamma'_{1}}{72m_{0}}\left(24C_{1}C_{2}-4C_{2}^{2}-3C_{5}^{2}\right)\nonumber \\
\nonumber \\
 &  & +\frac{\hbar^{4}\gamma_{2}}{3m_{0}}\left(3C_{1}^{2}-2C_{1}C_{2}+C_{2}^{2}-C_{5}^{2}\right)\nonumber \\
\nonumber \\
 &  & -\frac{\hbar^{4}\gamma_{3}}{24m_{0}}C_{5}\left(12C_{1}-2C_{2}+3C_{5}\right)
\end{eqnarray}

\begin{eqnarray}
\frac{\Omega_{5}}{k_{0}^{2}} & = & \frac{\hbar^{3}}{m_{\mathrm{e}}}C_{5}\nonumber \\
\nonumber \\
 &  & -\frac{\hbar^{4}\gamma'_{1}}{24m_{0}}C_{5}\left(24C_{1}-4C_{2}-3C_{5}\right)\nonumber \\
\nonumber \\
 &  & -\frac{\hbar^{4}\gamma_{2}}{12m_{0}}\left(12C_{1}C_{5}-2C_{2}C_{5}+3C_{5}^{2}\right)\nonumber \\
\nonumber \\
 &  & +\frac{\hbar^{4}\gamma_{3}}{24m_{0}}\left(72C_{1}^{2}-24C_{1}C_{2}-36C_{1}C_{5}\right.\nonumber \\
 &  & \qquad\quad\left.-16C_{2}^{2}-12C_{2}C_{5}+27C_{5}^{2}\right)
\end{eqnarray}
\end{subequations}


%

\end{document}